\documentclass[12pt]{article}
\usepackage{geometry}
\usepackage{a4}
\usepackage{graphicx}
\usepackage{epsf}
\usepackage{amsmath}
\usepackage{amssymb}
\usepackage{cite}
\usepackage{multirow,tabularx}
\newcommand{\be}{\begin{equation}}
\newcommand{\ee}{\end{equation}}

\newcommand{\Rmnum}[1]{\expandafter\@slowromancap\romannumeral #1@}
\newcommand{\bea}{\begin{eqnarray}}
\newcommand{\eea}{\end{eqnarray}}

\begin{document}
%%%%%%%%%%%%%%%%%%%%%%%%%%%%%%%
\def\A{{\mathbb{A}}}
\def\B{{\mathbb{B}}}
\def\C{{\mathbb{C}}}
\def\R{{\mathbb{R}}}
\def\s{{\mathbb{S}}}
\def\T{{\mathbb{T}}}
\def\Z{{\mathbb{Z}}}
\def\W{{\mathbb{W}}}
%%%%%%%%%%%%%%%%%%%%%%%%%%%%
\begin{titlepage}
\title{Thermodynamics, Phase Transition and Quasinormal modes with Weyl corrections}
\author{}
\date{% authors are dated
Subhash Mahapatra \footnote{subhmaha@imsc.res.in}  
\vskip1.6cm
The Institute of Mathematical Sciences, \\
Chennai 600113, India}
\maketitle
\abstract{ We study charged black holes in $D$ dimensional AdS space, in the
presence of four derivative Weyl correction. We obtain the black hole solution perturbatively up to first as well as second order in the Weyl coupling, and
show that first law of black hole thermodynamics is satisfied in all dimensions. We study its thermodynamic phase transition and then calculate the quasinormal frequencies of the massless scalar
field perturbation. We find that, here too, the quasinormal frequencies capture the essence of black hole phase transition.
Few subtleties near the second order critical point are discussed.

\noindent
}
\end{titlepage}

\section{Introduction}
Black holes are the simplest and yet the most mysterious objects of general relativity that are still far from being fully understood. Research in the theory of black holes has 
brought to light strong hints of a very deep and fundamental relationship between thermodynamics, gravitation and quantum theory. Although, full understanding of  quantum description
of black hole spacetime is still lacking, however a lot of progress has been made in understanding its thermodynamic properties using semi-classical approach. It has been long 
known that black holes are thermal objects, possessing entropy and temperature, and can undergo phase transitions like the typical thermodynamical systems in condensed matter physics 
\cite{HawkingPage}. For example, Schwarzschild black holes in asymptotically flat space are thermodynamically unstable, whereas asymptotically anti-de Sitter (AdS) ones can be in
thermal equilibrium with their own radiation. 

The recent developments in string theory and in particular the gauge/gravity duality \cite{Maldacena}-\cite{Witten} have attracted a lot of interest in black hole solutions in AdS spaces. 
The idea that the gauge/gravity duality can be potentially useful to understand strongly coupled field  theory at finite temperature have  generated a new and interesting field of application for
these solutions. Various thermodynamic properties of AdS black holes have been worked out and their interpretation in the dual boundary field theory side have been given.
For instance, Hawking-Page phase transition \cite{HawkingPage} from AdS-Schwarzschild black hole to thermal AdS is shown to be dual to the confinement/deconfinement transition 
in the boundary field theory \cite{Witten1}. The generalization to $U(1)$ charged AdS black hole was done in \cite{Chamblin}\cite{Chamblin1}, where a novel first order phase transition,
analogous to the Van der Waals liquid-gas system, between small and large black holes was found in the fixed charge ensemble. This was then extended into other charged AdS black holes,
in higher dimensions, with higher curvature gravity (see \cite{TDey}-\cite{Tsai}).

Apart from being theoretically important, there are a number of reasons to study charged black holes in AdS spaces.
From the gauge/gravity duality point of view, perhaps, the most
important reason is that they
provide a natural way to introduce chemical potential in the boundary system, which then can be used to model realistic strongly
coupled condensed matter systems through holography. In this regard, it will be of great importance from practical view point if more exact analytic charged black hole solutions can be obtained.
In particular, it might be more useful if charged black hole solutions can be obtained with additional parameters, as these additional parameters might provide non-trivial physics
in the boundary theory and can enhance our understanding of strongly coupled systems.

On the other hand, Quasinormal modes (QNMs)  describe the perturbations
in the surrounding geometry of a black hole. These are the solution of linearized wave equation about the 
black hole background, subject to the boundary condition that they are ingoing at the horizon. Since the perturbation can fall into a black hole and decay, the QNM
frequencies are complex. The real and imaginary part of the QNM frequencies describe oscillation and damping time of the perturbation respectively and are uniquely characterize 
by the black hole parameters. These QNMs are, therefore, believed to be the characteristic sounds of black holes and could be potentially detected in the gravitational wave detector
in the near future. For review on this topic see \cite{Kokkotas}-\cite{Zhidenko}.

Because of its astrophysical interest, QNMs in the asymptotically flat and di-Sitter black hole backgrounds have been extensively studied in the literature. However, with the
advent of gauge/gravity duality, the QNMs of AdS black holes have attracted a lot of attention of late \cite{Chan}. In \cite{Horowitz}, QNMs of the scalar field perturbation in the
AdS-Schwarzschild black hole background were calculated and physical interpretation of these QNMs, as the timescale to approach the thermal equilibrium, in dual boundary field theory 
was given. Similar analysis were then performed in RN-AdS background \cite{Wang}-\cite{Molina} as well as in higher derivative AdS black hole background \cite{Abdalla}.  

In the light of above discussion, it is a natural question to ask whether the dynamical perturbation in the black hole background can probe its different thermal phases. Indeed,
appealing investigations have recently started concerning probing of black hole phase transition using QNMs. This question
was first addressed in \cite{Koutsoumbas}, where they found that the QNMs of electromagnetic perturbation show distinct behavior in the MTZ black hole with scalar hair phase \cite{Martinez}
compared to the vacuum topological black hole phase. Signature of this phase transition in the QNMs 
was further established in \cite{Shen}, where QNMs of scalar perturbation were studied. QNMs of electromagnetic and gravitational perturbations in the backgrounds of
charged topological/AdS black holes were studied in \cite{Koutsoumbas1}, which
again showed contrasting behavior in different black hole phases. See \cite{Rao}\cite{Myung} for similar discussions on BTZ black hole phase transitions
and \cite{Maeda}\cite{He} for discussions on superconductor-metal/insulator phase transitions. In \cite{Liu}, Scalar QNMs in RN-AdS black hole background, near the first order transition line,
were studied and found that QNMs show distinct pattern in small and large black hole phases - both for the isobaric and isothermal phase transition processes. 
They clearly provided some evidence of the black hole phase transition in the QNMs behavior.
However, there still remains few 
subtle issues in probing ``small-large'' black hole phase transition through the scalar QNMs near the second order critical point, which 
were not discussed in \cite{Liu} and will be pointed out later in the text.

All the above mentioned work probing black hole phase transition through QNMs concerned with Einstein gravity with only two derivative terms in the action. However, it is also well known
that higher derivative interaction terms can arise in the quantum gravity corrections to
classical general relativity or in the low energy effective action of the string theory.
In holography, higher derivative interaction terms are known to produce non-trivial results in the bulk as well as in the field theory. For example, 
a non-trivial relation between the causality and viscosity bound for
Gauss-Bonnet gravity was found in \cite{Boer}, and in \cite{Soda}\cite{Dey0} it was shown that ratio of the energy gap to the critical temperature in
holographic superconductors
can be changed substantially by inclusion of higher derivative terms. Another important reason to consider higher derivative 
terms is that they provide extra tuneable coupling parameters in the system that might modify the physics non-trivially.
Therefore, one can naturally think of studying black hole thermodynamics and quasinormal modes, and its implications on the dual boundary theory, in the presence of higher 
derivative interaction terms. In this work we will undertake such an analysis.

We will consider a class of four derivative interaction terms, over and above the two derivative Einstein-Maxwell terms, that couple $U(1)$ gauge field to spacetime curvature.
In general a large number of such terms can be consistently added, however, the action can be considerably simplified by choosing particular linear combination of the
coupling constants \cite{Myers}\cite{Sachdev}. To be more precise here we will consider four derivative Weyl coupling correction, see the next section for more detail, to Einstein-Maxwell action. 
In recent years, the effects of this term have been thoroughly studied in literature. For example, corrections to the diffusion constant and conductivity due to Weyl coupling
constant $\gamma$ have been considered in \cite{Ritz}\cite{Sachdev1} and its effects in the process of holographic thermalization were studied in \cite{Dey1}
\footnote{See also \cite{Hofman}-\cite{Momeni}, for the discussion of Weyl coupling on other holographic studies.}.

The aim of this work is fourfold. The first will be to find charged AdS black hole solution with Weyl correction in general $D$ spacetime dimensions. Since exact solution is difficult to
find, here we will try to find the solution perturbatively by treating the Weyl coupling constant $\gamma$ as a small perturbative parameter. In four dimensions, linear order solution
in $\gamma$ was found in \cite{Dey2} and main purpose here is to find the solution for all spacetime dimensions. Our second aim is to find the black hole solution in higher 
order in $\gamma$ and show that the corrections due to higher order are small, and that the linear order analysis is trustable \footnote{To our knowledge only the leading order corrections due to
Weyl coupling
have been obtained in the literature and this is the first time that calculation at the nonleading order is shown.}. Here, we will explicitly calculate $\gamma^2$ corrected
black hole solution. Our third aim here is to calculate various thermodynamic variables and show that the first law of black hole thermodynamics is satisfied in Weyl corrected black holes
in all spacetime dimensions, not only at the leading order but also at the subleading order in $\gamma$. We will then study its thermodynamical properties in fixed charged ensemble
and will show the existence of familiar analogous Van der Waals like phase transition in it i.e., first order phase transition line terminating at the second order critical point. And finally our fourth aim here is to  study the quasinormal modes of 
the massless scalar field perturbation in Weyl corrected
black hole backgrounds. We will show that, here too, QNMs can be a good measure to probe the black hole phase transition. However, certain subtleties near the second order phase transition point will 
also be pointed out.

The paper is organized as follows : In section 2, we introduce the action and construct the black hole solution up to linear order
in $\gamma$. The black hole solution at next to leading order is shown in Appendix A. In section 3, we analyze the black hole thermodynamics in detail and 
explain the resemblance with Van der Waals-Maxwell liquid-gas system. In section 4, we numerically compute the quasinormal modes of the massless scalar field perturbation and relate
to the black hole phase transition of section 3. We conclude by summarizing our main results in section 5.

\section{Black hole solution with Weyl correction}
In this section, we first introduce the action and then construct a black hole solution in general $D$ dimensions by solving Einstein-Maxwell equations in the presence of negative
cosmological constant. We consider $D$ dimensional action in which gravity is coupled to $U(1)$ gauge field by two and four derivative interaction terms in the following way:
\begin{eqnarray}
&&\textit{S} = \frac{1}{16 \pi G_D} \int \mathrm{d^D}x \sqrt{-g} \ \ \bigl[R -2\Lambda
-\frac{1}{4}\textit{F}_{\mu\nu}\textit{F}^{\mu\nu}+ L^2 (a_1\textit{R}_{\mu \nu \rho \lambda}
\textit{F}^{\mu\nu}\textit{F}^{\rho\lambda} \nonumber\\
&&+a_2 R_{\mu\nu}\textit{F}^{\mu}\hspace{0.1mm} _{\rho}\textit{F}^{\nu\rho} + a_3 R F_{\mu\nu}F^{\mu\nu})\bigr] ,
\label{actiongeneral}
\end{eqnarray}
Here, $\varLambda$ is the negative cosmological constant related to AdS length scale $L$ as $\varLambda=-(D-1)(D-2)/2 L^{2}$, $F_{\mu\nu}$ is the field
strength tensor of $U(1)$ gauge field $A$ and $G_D$ is the Newton constant in D dimension. $a_1$, $a_2$ and $a_3$ are the  dimensionless coupling constants of four derivative
interaction terms between gauge field and spacetime curvature tensor.

The above action can be greatly simplified if we consider a certain linear combination of the four derivative coupling constants. Following \cite{Myers}\cite{Sachdev} and as elaborated in
\cite{Dey1}, we can write the action by choosing a specific combination of $a_1$, $a_2$ and $a_3$ as 
\begin{equation}
\textit{S} = \frac{1}{16 \pi G_D}\int \mathrm{d^D}x \sqrt{-g} \ \ \bigl[R+\frac{(D-1)(D-2)}{L^{2}}
-\frac{1}{4}\textit{F}_{\mu\nu}\textit{F}^{\mu\nu}+\gamma L^2 \textit{C}_{\mu \nu\rho\lambda}
\textit{F}^{\mu\nu}\textit{F}^{\rho\lambda}\bigr] ,
\label{action}
\end{equation}
where, $\gamma$ represents the effective coupling for four derivative interaction terms and $\textit{C}_{\mu \nu\rho\lambda}$ is the Weyl tensor which in $D$ spacetime dimensions is given by, 
\begin{eqnarray}
 \textit{C}_{\mu \nu\rho\lambda}=\textit{R}_{\mu \nu\rho\lambda}+{1\over (D-2)}(g_{\mu \lambda}R_{\rho \nu}+g_{\nu \rho} 
 R_{\mu \lambda}-g_{\mu \rho}R_{\lambda \nu}-g_{\nu \lambda}R_{\rho \mu}) \nonumber \\
 +{1\over (D-1)(D-2)}(g_{\mu \rho}g_{\nu \lambda}-
 g_{\mu \lambda}g_{\rho \nu})R .
\label{Weyl tensor}
\end{eqnarray}
In this paper we will refer $ \gamma$ as the `Weyl coupling'.

Now by varying the action, we find the equation of motion (EOM) for the gauge field as
\begin{eqnarray}
 \nabla_{\mu}(F^{\mu\lambda}-4\gamma L^2 C^{\mu\nu\rho\lambda}F_{\nu\rho})=0 .
 \label{MaxwellEOM}
\end{eqnarray}

On the other hand, the Einstein equation reads
\begin{eqnarray}
R_{\mu\nu}-{1\over2}g_{\mu\nu}R-{(D-1) (D-2)\over 2 L^2}g_{\mu\nu}-T_{\mu\nu}=0 \,,
\label{EinsteinEOM}
\end{eqnarray}
Where, the energy-momentum tensor $T_{\mu\nu}$ is given by,
\begin{eqnarray}
 && T_{\mu\nu}={1\over 2}\bigl(g^{\alpha\beta}F_{\mu\alpha}F_{\nu\beta}-{1\over4}g_{\mu\nu}F^2\bigr)+
 {\gamma L^2}\bigl[{1 \over2}g_{\mu\nu}C_{\delta\sigma\rho\lambda}F^{\delta\sigma}F^{\rho\lambda}-3 g_{\delta\mu}
 R_{\nu\sigma\rho\lambda}F^{\delta\sigma}F^{\rho\lambda} \nonumber \\
 &&-\nabla_{\beta}\nabla_{\rho}(F_{\mu}\hspace{0.1mm}^{\beta}F_{\nu}\hspace{0.1mm}^{\rho})-\nabla_{\rho}\nabla_{\beta}(F_{\mu}\hspace{0.1mm}^{\beta}F_{\nu}\hspace{0.1mm}^{\rho})
 +{2 \over D-2} \biggl(\nabla^{2}
 (F_{\mu}\hspace{0.1mm}^{\rho}F_{\nu\rho})+ g_{\mu\nu}\nabla_{\sigma}\nabla_{\delta}(F^{\delta}
 \hspace{0.1mm}_{\rho}F^{\sigma\rho}) \nonumber \\ &&
 -\nabla_{\rho}\nabla_{\mu}(F_{\nu\beta}F^{\rho\beta}) -\nabla_{\rho}\nabla_{\nu}(F_{\mu\beta}F^{\rho\beta}) \biggr)
  + {4 \over D-2}\biggl(R_{\mu\rho}F^{\rho\beta}F_{\nu\beta} + R_{\nu\rho}F^{\rho\beta}F_{\mu\beta} + R_{\rho\beta}F_{\mu}\hspace{0.1mm}^{\beta}F_{\nu}\hspace{0.1mm}^{\rho}  \biggr) \nonumber \\ &&
  - {2 \over(D-1)(D-2)} \biggl(R_{\mu\nu}F^2 +2RF_{\mu\rho}F_{\nu}\hspace{0.1mm}^{\rho} + g_{\mu\nu}\nabla^{2}F^2 
  -{1\over2}(\nabla_{\mu}\nabla_{\nu}+\nabla_{\nu}\nabla_{\mu})F^2
  \biggr)
\label{EnergyMomentumTensor}
\end{eqnarray}
In the above equation, notations $F^2=F_{\alpha\beta}F^{\alpha\beta}$ and $\nabla^2=\nabla_{\rho}\nabla^{\rho}$ have been used. 

\vspace{0.5cm}
Now, we will construct a black hole solution in $D$ dimensions by solving eqs. (\ref{MaxwellEOM}) and (\ref{EinsteinEOM}) simultaneously. A solution in four spacetime dimensions has been constructed 
\cite{Dey2}, however, here we will  construct a black hole solution in general $D$ dimension.
We will concentrate on the horizon with spherical topology, as the horizon with planner topology can be straightforwardly generalized. In order to solve 
eqs. (\ref{MaxwellEOM}) and (\ref{EinsteinEOM}), we consider the following ansatz for the gauge field
and metric:
\begin{equation}
 A = \phi(r)dt \,.
 \label{A ansatz}
\end{equation}
\begin{eqnarray}
ds^2&=&-  f(r) e^{-2\chi(r)}  dt^2+{1\over f(r)}dr^2 +r^2 d\Omega^2_{D-2} \,,
\label{metric ansatz}
\end{eqnarray}
where, $d\Omega^2_{D-2}$ is the metric on $D-2$ dimensional sphere. 

\vspace{0.5cm}
The presence of $\gamma$ in Einstein and Maxwell equations make them extremely difficult to solve exactly. Hence, we will try to solve them perturbatively
in linear order in $\gamma$. Form a string theory perspective, these higher derivative interaction terms are expected to arise as quantum corrections to two derivative action, and therefore our 
assumption seems reasonable. In order to further justify the perturbative analysis, in Appendix A, we will solve the Einstein and Maxwell equations in next to leading order in $\gamma$. 
There we will explicitly show that, for small values of $\gamma$, the $\gamma^2$ corrections are small and that the leading order analysis is trustworthy.

With this assumption in mind, we now proceed to find gauge field and metric in leading order in $\gamma$. For this purpose, we consider the following ansatz for 
$\phi(r)$, $\chi(r)$ and $f(r)$
\begin{eqnarray}
\phi(r) &=& \phi_0(r)+ \gamma \phi_1(r),  \nonumber\\
\chi(r) &=& \chi_0(r) + \gamma \chi_1(r)\, \nonumber\\
f(r)&=&f_0(r)\bigl(1+\gamma \mathcal{F}_{1}(r)\bigr)\
\label{perturbationansatz}
\end{eqnarray}
where $\phi_0(r)$, $\chi_0(r)$ and $f_0(r)$ are the zeroth order solutions representing a Reissner-Nordstr\"{o}m black hole solution in AdS space
with,
\begin{eqnarray}
\phi_{0}(r) &=&  \sqrt{{2 (D-2) \over (D-3)}}q \bigl({1\over r_{h}^{D-3}}-{1\over r^{D-3}}\bigr), \nonumber\\
\chi_{0}(r) &=& 0 \,, \nonumber\\ 
f_{0}(r) &=& 1-\frac{m}{r^{D-3}}+\frac{q^2}{r^{2D-6}}+\frac{r^2}{L^2} \,,
\label{zeroth order soln}
\end{eqnarray}
here $r_h$ is the radius of the event horizon and $m$ is an integration constant related to the ADM mass ($M$) of the black hole, to be examined in detail in the next section. The other
integration constant $q$ is related to the total charge $Q$ of the black hole.

In eq. (\ref{perturbationansatz}): $\phi_1(r)$, $\chi_1(r)$ and $\mathcal{F}_{1}(r)$ are the linear order corrections. Expressions for these quantities can be obtained by solving eqs.
(\ref{MaxwellEOM}) and (\ref{EinsteinEOM}) while keeping the terms up to linear order in $\gamma$. We find the solution of these correction terms as,

\begin{eqnarray}
\chi_{1}(r) &=& k_2 -2 {(D-3)^3 \over (D-1)} {L^2 q^2 \over r^{2D-4}}
\label{chi1EOM}
\end{eqnarray}
\begin{eqnarray}
\phi_{1}(r) = {2 (7D^2-30D+31) \sqrt{2(D-2)(D-3)^3}\over(3D-7)(D-1)}{L^2 q^3\over r^{3D-7}}  \nonumber\\
-2 \sqrt{2(D-2)(D-3)^3} {L^2 m q \over r^{2D-4}} -{\sqrt{2 (D-2)\over (D-3)}}{k_4 q \over r^{D-3}} + k_3
\label{phi1EOM}
\end{eqnarray}
\begin{eqnarray}
\mathcal{F}_{1}(r) = {1\over f_{0}(r)} \biggl[{-8(D-2)(2D-5)(D-3)^2 \over(3D-7)(D-1)}{L^2 q^4\over r^{4D-10}} - {8(D-3)^3\over(D-1)}{L^2 q^2\over r^{2D-4}} \nonumber\\
+ {2(3D-7)(D-3)^2 \over (D-1)}{L^2 m q^2 \over r^{3D-7}} - {8(D-2)(D-3)^2 \over(D-1)}{q^2\over r^{2D-6}} \nonumber\\
+ {2k_4 q^2\over r^{2D-6}} + {k_1 r_{h}^{D-1}\over L^2 r^{D-3}} + {2 q^2 k_2 \over r^{2D-6}}\biggr] 
\label{F1EOM}
\end{eqnarray}
where $k_1$, $k_2$, $k_3$ and $k_4$ are dimensionless integration constants. These constants can be determined by imposing certain constrains \cite{Myers}. For example, near the 
asymptotic boundary $r\rightarrow\infty$, the metric behave as
\begin{eqnarray}
 ds^2 \vert_{r\rightarrow \infty}=- (f e^{-2\chi})_\infty dt^2+r^2 (d\theta^2+\sin ^2  \theta \ d\phi^2) .
   \label{CFT metric}
\end{eqnarray}
where, $(f e^{-2\chi})_\infty=\lim_{r\to\infty}f(r) e^{-2\chi(r)}$. The metric in eq. (\ref{CFT metric}) represents the background metric where the dual boundary theory lives. In order to fix
the  speed of light to unity in the boundary theory, we demand that $(f e^{-2\chi})_\infty={r^2 \over L^2}$, which in turn gives $k_2=0$.

Similarly, in order to determine $k_4$ we impose the constraint that the charge density $q$ remains unchanged. Note that, we can recast the Maxwell equation 
(\ref{MaxwellEOM}) in the form $\nabla_{\mu}X^{\mu\lambda}=0$, where $X^{\mu\lambda}$ is an antisymmetric tensor. Hence, the 
dual of $(*X)_{\theta_1...\theta_{D-2}}$, where $\theta_1...\theta_{D-2}$ are the coordinates of $D-2$ sphere, is a constant and it is convenient to choose this constant to be the 
fixed charge density $q$, 
i.e., $(*X)_{\theta_1...\theta_{D-2}}=q$. Since the quantity $(*X)_{\theta_1...\theta_{D-2}}$ does not depend on
radial coordinate $r$, we demand 
\begin{eqnarray}
 \lim_{r\rightarrow\infty}\left(*X\right)_{\theta_1...\theta_{D-2}}=q .
 \label{constraint1a}
\end{eqnarray}
On the other hand, computation of this quantity in the $r\rightarrow\infty$ limit gives,
\begin{eqnarray}
 \lim_{r\rightarrow\infty}\left(*X\right)_{\theta_1...\theta_{D-2}}&=&\lim_{r\rightarrow\infty}\bigl[r^{D-2} \omega_{d-2} e^{\chi(r)}
 \left(F_{r t}-8\gamma L^2 C_{r t}\hspace{0.1mm}^{r t}F_{r t}\right)\bigr] \nonumber\\
 &=&\bigl(1 + \gamma k_4 \bigr)q .
\label{constraint1b}
\end{eqnarray}
where $\omega_{D-2}$ is the volume of the unit 
$(D-2)$ sphere. A comparison of equations (\ref{constraint1a}) and (\ref{constraint1b}) yields, $k_4=0$.

The constant $k_3$ can be determined by demanding $\phi(r)$ to be vanished at the horizon. This is imposed to have a well defined one-form for the
gauge field. After imposing $\phi(r_h)=0$, we get
\begin{eqnarray}
 k_3= - {2 (7D^2-30D+31) \sqrt{2(D-2)(D-3)^3}\over(3D-7)(D-1)}{L^2 q^3\over r_{h}^{3D-7}}  \nonumber\\
+2 \sqrt{2(D-2)(D-3)^3} {L^2 m q \over r_{h}^{2d-4}} 
 \label{K3EOM}
\end{eqnarray}
 
Similarly, the remaining constant $k_1$ can be determined by fixing the position of the event horizon, i.e. $f_0(r) \mathcal{F}_{1}(r) |_{r=r_h}=0$. We get
\begin{eqnarray}
k_{1} = {L^2\over r_{h}^2} \biggl[{ 8(D-2)(2D-5)(D-3)^2 \over(3D-7)(D-1)}{L^2 q^2\over r_{h}^{4D-10}} + {8(D-3)^3\over(D-1)}{L^2 q^2\over r_{h}^{2D-4}} \nonumber\\
- {2(3D-7)(D-3)^2 \over (D-1)}{L^2 m q^2 \over r_{h}^{3D-7}} + {8(D-2)(D-3)^2 \over(D-1)}{q^2\over r_{h}^{2D-6}}  \biggr] 
\label{K1EOM}
\end{eqnarray}
Now we have determined all the constants. Below we write down the final expressions for $\phi_1(r)$, $\chi_1(r)$ and $\mathcal{F}_{1}(r)$ for completeness,
\begin{eqnarray}
\chi_{1}(r) &=&-2 {(D-3)^3 \over (D-1)} {L^2 q^2 \over r^{2D-4}}
\label{chi1EOM1}
\end{eqnarray}
\begin{eqnarray}
\phi_{1}(r) = 2\sqrt{2(D-2)(D-3)^3} \biggl[{(7D^2-30D+31) \over(3D-7)(D-1)}\biggl(\frac{L^2 q^3}{r^{3D-7}}-\frac{L^2 q^3}{r_{h}^{3D-7}}\biggr)  \nonumber\\
- L^2 m q \biggl({1 \over r^{2D-4}}-{1 \over r_{h}^{2D-4}} \biggr) \biggr]
\label{phi1EOM1}
\end{eqnarray}
\begin{eqnarray}
\mathcal{F}_{1}(r) = {(D-3)^2 \over(3D-7)(D-1)f_{0}(r)} \biggl[-{8(D-2)(2D-5)L^2 q^4\over r^{4D-10}} + { 2(3D-7)^2 L^2 m q^2 \over r^{3D-7}}
\nonumber\\ - {8(D-3)(3D-7)L^2 q^2\over r^{2D-4}} 
 - { 8(D-2)(3D-7) q^2\over r^{2D-6}} + {2(5D^2-22D+25)m\over r^{D-3}} \nonumber\\ 
- {2(D-3)^2 L^2m^2 \over r^{D-3}r_{h}^{D-1}} -{8(2D-5)(D-3)r_{h}^{D-3}\over r^{D-3}} - {8(D^2-7D+11)L^2 r_{h}^{D-5}\over r^{D-3}} \nonumber\\ 
+{2 (5D^2-34D+53)L^2 m \over r^{D-3}r_{h}^2} - {8 (D-2)^2 r_{h}^{D-1} \over L^2 r^{D-3} }
    \biggr] 
\label{F1EOM1}
\end{eqnarray}
The Hawking temperature of the black hole is given by, 
\begin{eqnarray}
T={\kappa \over 2\pi}=\frac{\biggl(f_{0}(r)(1+\gamma \mathcal{F}_{1}(r)) \biggr)' e^{-\gamma\chi_{1}(r)}}{4\pi}\Big \vert_{r=r_h} 
\label{Hawking Temp}
\end{eqnarray}
where $\kappa$ is the surface gravity.

Before proceeding, a word about the next to leading order corrections to metric and gauge field is in order. Above we have explicitly shown how to calculate linear order corrections 
to metric and gauge field, due to Weyl coupling, perturbatively by solving Einstein and Maxwell equations at linear order in $\gamma$. This perturbative analysis can in principle be 
generalized to calculate higher order corrections. However, this will amount to solve Einstein and Maxwell equations at higher order in $\gamma$ which is a daunting task. 
Nevertheless, using the perturbative analysis we are able to solve them at next to leading order. In Appendix A, we have explicitly calculated metric and gauge field corrections at order
$\gamma^2$. There, we have shown the calculation only for four dimensions but our analysis can be straightforwardly extended to higher dimensions. In subsequent sections, when we will 
discuss black hole thermodynamics and QNMs, we will consider only those values of $\gamma$ for which $\gamma^2$ corrections are small and do not change the leading order analysis significantly. 
Small corrections at $\gamma^2$ order will further justify and substantiate our results at the leading order.

\section{\textbf{Black hole thermodynamics with Weyl correction}}

We now proceed to study thermodynamics of the linear order D dimensional Weyl corrected black hole geometry obtained in the last section. For this purpose,
we first need to construct the on-shell action. A straightforward calculation in D dimensions yields,
\begin{eqnarray}
S_{on-shell}=\frac{1}{16 \pi G_D} \int d^D x \sqrt{-g} \biggl[\frac{F_{\mu\nu}F^{\mu\nu}}{2(D-2)} + \frac{2(D-1)}{L^2} - \frac{4\gamma C_{\mu\nu\rho\lambda} F^{\mu\nu}F^{\rho\lambda}}{(D-2)} 
 \biggr]
\end{eqnarray}
\begin{eqnarray}
S_{on-shell}
\xrightarrow{r\rightarrow\infty}\frac{\beta \omega_{D-2}}{8 \pi G_D} \biggl(r^{D-1} 
-\frac{q^2}{r_{h}^{D-3}}-r_{h}^{D-1}\biggr) \nonumber\\ +
\frac{\beta \omega_{D-2}(D-3)^2 \gamma q^2}{4 \pi G_D}\biggl(\frac{1}{r_{h}^{D-3}}  
-\frac{(D-3)^2q^2}{(3D-7)(D-1) r_{h}^{3D-7}}\biggr) 
\label{onshellaction}
\end{eqnarray}
Here, $\beta=\frac {1}{T}$ is the inverse temperature  and we have set AdS radius $L=1$. We will use this unit throughout the paper. It can be seen from the 1st term of
eq. (\ref{onshellaction}) that the on-shell action is diverging in the $r\rightarrow\infty$ limit. In order to remove this divergent part we add Gibbons-Hawking (GW) and 
counter term(CT) at the boundary.  The standard forms of these quantities are
\begin{eqnarray}
S_{GH}=-\frac{1}{8 \pi G_D} \int d^{D-1} x \sqrt{-\sigma} \Theta
\end{eqnarray}
\begin{eqnarray}
S_{CT}=\frac{1}{16 \pi G_D} \int d^{D-1} x \sqrt{-\sigma} \biggl[2(D-2) 
+\frac{\mathcal{R}_{D-1}}{D-3} \nonumber\\
+ \frac{1}{(D-5)(D-3)^2}\biggl( \mathcal{R}_{ab}\mathcal{R}^{ab} -\frac{(D-1)\mathcal{R}_{D-1}^2}{4 (D-2)}  \biggr)+... \biggr] 
\end{eqnarray}
Here, $\sigma$ is the induced metric on the boundary and $\Theta$ is the trace of the extrinsic curvature. $\mathcal{R}_{D-1}$ is the Ricci scalar constructed from $D-1$ dimensional 
boundary metric and $\mathcal{R}_{ab}$ is the corresponding Ricci Tensor. With these additional terms the total action $S_{Total}= S_{on-shell} + S_{GH} + S_{CT }$ is now divergence-less, 
which is given by,
\begin{eqnarray}
S_{Total} =\frac{\omega_{D-2}\beta }{16\pi G_D } \biggl( r_{h}^{D-3} -r_{h}^{D-1} -\frac{q^2}{r_{h}^{D-3}}  \biggr) +\frac{\omega_{D-2} \beta \gamma q^2 (D-3)^2}{8 \pi G_D } 
\biggl( \frac{1}{r_{h}^{D-3}} \nonumber\\  - \frac{D-5}{(D-1) r_{h}^{D-1}}-\frac{q^2 (D-3)^2}{(3D-7)(D-1) r_{h}^{3D-7}}\biggr) 
\label{stotal}
\end{eqnarray}
For odd number of dimensions there also appears a constant Casimir energy term which we have neglected here, as this term is immaterial for the discussion of black hole thermodynamics.
We will identify $S_{Total}$ times the temperature, using the gauge/gravity duality, as the free energy of the system.

\subsection{\textbf{First law of Weyl corrected black hole thermodynamics}}
In this section, we will establish the first law of black hole thermodynamics for linear order Weyl corrected black hole geometry constructed in the previous section. In \cite{Dey2},
we had verified first law  in four dimensions and the main aim of this section is to show that this law is satisfied in all dimensions. In Appendix A, we will go a step further and show that 
the first law of thermodynamics is also satisfied in next to leading order. In order to establish the first law, we first need to calculate the mass
of the black hole. For this purpose, we use Ashtekar-Magnon-Das (AMD) prescription which gives a mechanism to calculate conserved quantity $\mathcal{C}[K]$ in an asymptotically AdS spacetime 
associated with a  Killing field $K$ as 
\footnote{For detail analysis, we refer readers to \cite{AshtekarDas}.}
\begin{equation}
\mathcal{C}[K]=\frac{1}{8 \pi (D-3)G_D} \oint \tilde{\epsilon}^{\mu}_{ \ \nu}  K^{\nu} d\tilde{\varSigma}^{\mu}
\label{charge}
\end{equation}
Here $K^{\nu}$ is the conformal killing vector field,  $\tilde{\epsilon}^{\mu}_{ \ \nu}=\Omega^{D-3} \tilde{n}^\rho \tilde{n}^\sigma \tilde{C}^{\mu}_{\ \ \rho \nu \sigma}$, 
$\tilde{C}^{\mu}_{\ \ \rho \nu \sigma}$ is the Weyl tensor constructed from $\tilde{ds^2}=\Omega^2 ds^2 $ with $\Omega=1/r$ and
$\tilde{n}^\rho$ is the unit normal vector. 
$d\tilde{\varSigma}^{\mu}$ is the $D-2$ dimensional area element of the transverse section of the AdS  boundary. For a
timelike killing vector, we get the following form for the conserved mass
\begin{equation}
\mathcal{C}[K]=M=\frac{\omega_{D-2}}{8 \pi (D-3) G_D} \Omega^{3-D} (\tilde{n}^{\Omega})^{2}\tilde{C}^{t}_{\ \ \Omega t \Omega} 
\end{equation}
substituting the form of $\tilde{C}^{t}_{\ \ \Omega t \Omega}$ and converting back to $r=1/\Omega$ coordinate, we get the following form for the black hole mass M,
\begin{equation}
M=-\frac{\omega_{D-2}}{8 \pi G_D} \frac{r^{D-1}}{(D-1)} \biggl(\frac{g_{rr} g_{tt}''}{2 g_{tt}} - \frac{g_{rr} g_{tt}'^2}{4 g_{tt}^2} + \frac{g_{rr}' g_{tt}'}{4 g_{tt}} 
-\frac{g_{rr} g_{tt}'}{2 r g_{tt}} -\frac{g_{rr}'}{2 r} -\frac{(1-g_{rr})}{r^2}\biggr)
\end{equation}
where $$g_{tt}=f_{0}(r)(1+ \gamma \mathcal{F}_{1}(r))e^{-2\gamma \chi_{1}(r)}, \ \ g_{rr}=f_{0}(r)(1+\gamma \mathcal{F}_{1}(r))$$ and  $'$ denotes the derivative with respect to $r$.
Now, substituting the forms of $f_{0}(r)$, $\mathcal{F}_{1}(r)$ and $\chi_{1}(r)$, we get
\begin{eqnarray}
&& M=\frac{(D-2)\omega_{D-2}}{16 \pi G_D} \biggl(\frac{q^2}{r_{h}^{D-3}} +r_{h}^{D-3} +r_{h}^{D-1}\biggr) \nonumber\\
&& +\frac{\omega_{D-2} \gamma q^2 (D-2)(D-3)^2}{8 \pi G_D} \biggl(\frac{(5-D)}{(D-1) r_{h}^{D-1}} 
-\frac{1}{r_{h}^{D-3}} +\frac{q^2 (D-3)^2}{(3D-7)(D-1) r_{h}^{3D-7}}\biggr) \nonumber\\
\label{AMDmass}
\end{eqnarray}
From eq. (\ref{AMDmass}), we note that the expression of $M$ in $\gamma=0$ limit reduces to that of standard ADM mass of RN-AdS black hole in D dimensions. Now, we proceed to calculate the 
entropy of the black hole. Since our gravity theory is four derivative, therefore, we need to calculate Wald entropy \cite{Wald} which is given as
\begin{eqnarray}
S_{Wald}=-2\pi \int d^{D-2} x \sqrt h \frac{\partial \mathcal{L}}{\partial R_{\mu\nu\rho\lambda}} \varepsilon_{\mu\nu} \varepsilon_{\rho\lambda}
\end{eqnarray}
Here $h$ is the determinant of the matrix of $D-2$ dimensional killing horizon, $\mathcal{L}$ is the Lagrangian and $\epsilon_{\mu\nu}$ is the binormal killing vector 
normalized by $\varepsilon^{\mu\nu}\varepsilon_{\mu\nu}=-2$. For our system (eq. (\ref{action})), we get
\begin{eqnarray}
&& S_{Wald}=-\frac{1}{8 G_D} \int d^{D-2}x \sqrt h \bigg[\biggl(1+\frac{2 \gamma F_{\mu\nu}F^{\mu\nu}}{(D-1)(D-2)}\biggr)g^{\mu\rho}g^{\nu\lambda}\varepsilon_{\mu\nu}\varepsilon_{\rho\lambda}
\nonumber\\ && \ \ \ \ \ \ \ \ \ \ \ \ \ \ \ \ \ \ \ \ \ \ \ \   + \gamma F^{\mu\nu}F^{\rho\lambda}\varepsilon_{\mu\nu}\varepsilon_{\rho\lambda} 
+ \frac{4\gamma}{(D-2)} g^{\mu\nu}F^{\sigma\rho}F_{\rho}^{\ \lambda}\varepsilon_{\sigma\mu}\varepsilon_{\lambda\nu} \biggr] \nonumber\\
&& \ \ \ \ \ \ \ \ \ \ =\frac{\omega_{D-2} r_{h}^{D-2}}{4 G_D}-\frac{\omega_{D-2} \gamma q^2}{r_{h}^{D-2} G_D}\frac{(D-2)(D-3)^2}{(D-1)}
\label{WaldEntropy}
\end{eqnarray}
For completeness, let us also note the expressions of conserved charge $(Q)$ and potential $(\Phi)$
\begin{eqnarray}
&& Q=\sqrt{\frac{(D-2)(D-3)}{2}}\frac{q \omega_{D-2}}{8 \pi G_{D}} \nonumber\\
&& \Phi= \sqrt{{2 (D-2) \over (D-3)}}{q\over r_{h}^{D-3}} + 2\gamma \sqrt{2(D-2)(D-3)^3} \biggl[{L^2 mq\over r_{h}^{2D-4}} \nonumber\\ 
&& \ \ \ \ \ \ \ \ \ \ \ \ \ \ \ \ \ \ \ \ \ \ \  \ \ \ \ \ \ \ \ \ \ \ \ \ \  - {(7D^2-30D+31) \over(3D-7)(D-1)} \frac{L^2 q^3}{r_{h}^{3D-7}} \biggr]
\label{conservedcharge}
\end{eqnarray}
Finally, we get the Gibbs free energy as,
\begin{eqnarray}
&&G=M-T S_{Wald} -Q \Phi =\frac{\omega_{D-2}}{16\pi G_D } \biggl( r_{h}^{D-3} -r_{h}^{D-1} -\frac{q^2}{r_{h}^{D-3}}  \biggr) \nonumber\\ 
&& +\frac{\omega_{D-2} \gamma q^2 (D-3)^2}{8 \pi G_D } 
\biggl( \frac{1}{r_{h}^{D-3}} 
  - \frac{(D-5)}{(D-1) r_{h}^{D-1}}-\frac{q^2 (D-3)^2}{(3D-7)(D-1) r_{h}^{3D-7}}\biggr) 
\label{1stlaw}
\end{eqnarray}
which is nothing but $G=S_{Total}/\beta$. This essentially implies that the first law black hole thermodynamics is satisfied in all dimensions for the linear order Weyl corrected geometry. 

\subsection{Thermodynamics}
After establishing the 1st law of black hole thermodynamics for Weyl corrected black hole geometry now, in this section, we proceed to discuss its thermodynamic properties.
Here, we will mostly focus on the fixed charged ensemble, as this is the ensemble which exhibits many interesting features. Similar study for the fixed potential ensemble 
can be straightforwardly generalized, but we will not discuss it here. 
\\
Let us first note the expression of Helmholtz free energy at the leading order in $\gamma$:
\begin{eqnarray}
&& F= G  + Q \Phi =\frac{\omega_{D-2}}{16\pi G_D } \biggl( r_{h}^{D-3} -r_{h}^{D-1} + \frac{(2D-5)q^2}{r_{h}^{D-3}}  \biggr) 
 +
\biggl( \frac{(2D-3)}{r_{h}^{D-3}} \nonumber\\
&& +\frac{(2D^2-7D+9)}{(D-1)r_{h}^{D-1}} -\frac{(D-3)(8D^2-31D+29)q^2}{(3D-7)(D-1)r_{h}^{3D-7}}\biggr) \frac{\omega_{D-2} \gamma q^2 (D-3)^2 }{8 \pi G_D } 
\label{Helmholtz}
\end{eqnarray}
Below, for computational purposes, we set $\omega_{D-2}=1$. 

%%%%%%%%%%%%%%%%%%%%%%%%%%%%%%
\begin{figure}[t!]
\begin{minipage}[b]{0.5\linewidth}
\centering
\includegraphics[width=2.8in,height=2.3in]{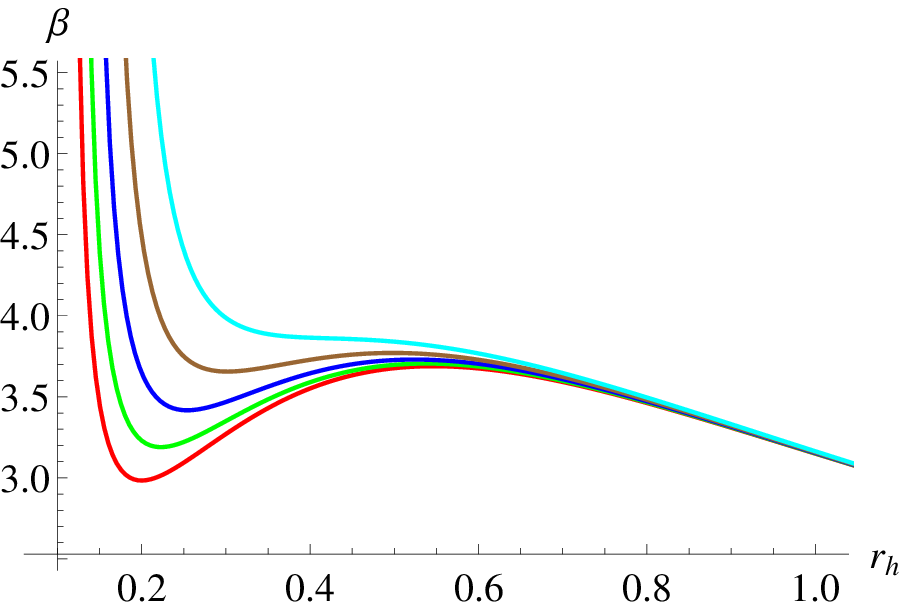}
\caption{ \small $\beta$ as a function of horizon radius for $\gamma=0.001$ in four dimensions. Red, green, blue, brown and cyan curves correspond to 
${1\over10}$, ${1\over 9}$, ${1\over 8}$, ${1\over 7}$ and ${1\over6}$ respectively.}
\label{4DbetaVsRhgamma1by1000vsq}
\end{minipage}
\hspace{0.4cm}
\begin{minipage}[b]{0.5\linewidth}
\centering
\includegraphics[width=2.8in,height=2.3in]{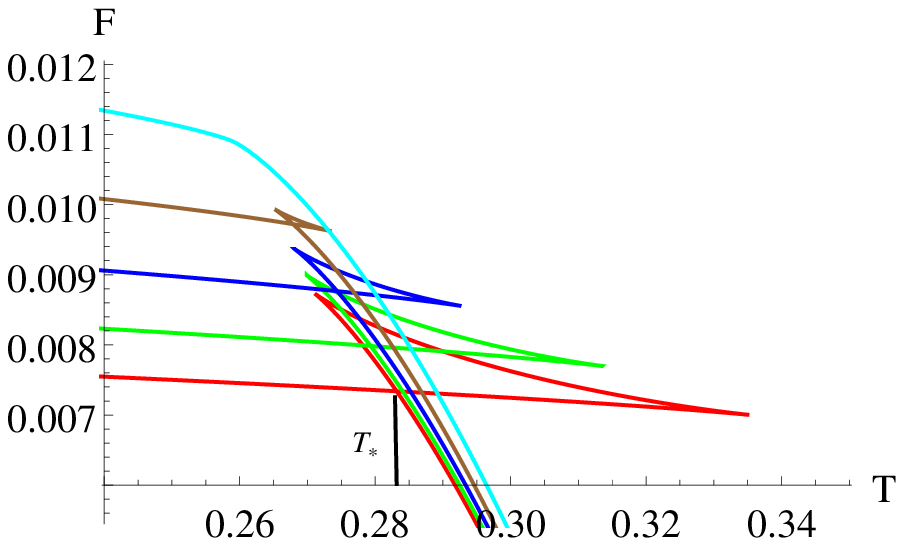}
\caption{\small Free energy as a function of temperature for $\gamma=0.001$ in four dimensions. Red, green, blue, brown and cyan curves correspond to 
${1\over10}$, ${1\over 9}$, ${1\over 8}$, ${1\over 7}$ and ${1\over6}$
respectively.}
\label{4DFVsTgamma1by1000vsq}
\end{minipage}
\end{figure}
%%%%%%%%%%%%%%%%%%%%%%%%%%%%%%
Based on eqs. (\ref{Hawking Temp}) and (\ref{Helmholtz}), we will discuss the phase structure of Weyl corrected black hole geometries. Let us
first consider the thermodynamics in $D=4$ dimension.
The influence of
charge on the phase structure for a fixed $\gamma=0.001$ is shown in fig.(\ref{4DbetaVsRhgamma1by1000vsq}), where variation of inverse Hawking temperature with
respect to horizon radius is plotted. Here red, green, blue, brown and cyan curves 
correspond to $q$ = ${1\over 10}$, ${1 \over9}$, ${1\over 8}$, ${1\over 7}$  and ${1\over6}$ respectively. Apparently, the phase structure is similar to that of the 
RN-AdS black hole. For small value of q, say  $q=1/10$, there exist three black hole branches. 
Two of these branches, which occur at small and large black hole radii, have negative slope and are stable. Where as the other branch, which has positive slope, corresponds to intermediate
black hole and is unstable. 

The presence of two stable black hole branches which are connected by a unstable intermediate branch suggests a possible first order phase transition from a small black hole to a large black hole as
we increase the temperature. This is indeed the case as can be seen from fig.(\ref{4DFVsTgamma1by1000vsq}), where free energy as a function of temperature is plotted. 
Here, we have used the same coding as in fig.(\ref{4DbetaVsRhgamma1by1000vsq}).  The swallow tail like behavior of the free energy, which is a characteristic feature of first order
phase transition, is apparent for $q=1/10$. A black vertical line, whose upper point terminates at the kink of the swallow tail, indicates the critical temperature $(T_*)$
where free energy of the large black hole becomes smaller than the small black hole. For $\gamma=0.001$ and $q=1/10$, we find $T_*\simeq0.283$.

However, the scenario changes as we increase the values of $q$. With increase in $q$, the structure of swallow tail start decreasing in size and completely disappear above a certain 
critical charge $q_c$. At $q_c$, the small and the large black hole merge into one and forms a single black hole which is stable  at all the temperatures. This can be seen from cyan 
curve of figs.(\ref{4DbetaVsRhgamma1by1000vsq}) and (\ref{4DFVsTgamma1by1000vsq}) where $q=1/6$ is considered. We see that black holes
endowed with different $q's$ have different
free energy behavior. For small $q$, there is
an unstable black hole interpolating between stable small and stable large black hole whereas for larger $q's$ this situation is absent.

Also, We find that the specific heat at constant charge defined by $$C_q=T(\partial S_{Wald} / \partial T)$$
diverges at $q_c$. The divergence of $C_q$ implies that the phase transition at $q_c$ is of second order. Therefore, our whole analysis indicate a line of first order phase
transition between black holes of different sizes which terminates at the second order critical point. This is analogous to the classic Van der Waals liquid-gas 
phase transition in black holes which was first discovered in \cite{Chamblin}. In order to find the critical exponent of $C_q$, we perform a series expansion of 
$T(S_{Wald})$ at $q=q_c$. After a little bit of algebra, we find that
\begin{equation}
 C_{q}(T)=T({\partial S_{Wald}\over \partial T}) \propto (T-T_c)^{-{2\over 3}} .
 \label{BHSpecificHeat}
\end{equation}
which shows that the critical exponent of the specific heat in our Weyl corrected black hole geometry is same as in RN-AdS black hole case.
\\
\\
Now a  word regarding the dependence of $q_c$ on $\gamma$ is in order. In order to find $q_c$, we note that $q_c$ defines 
an inflection point in the $\beta-r_h$ plane. Therefore, at this point, following two conditions must be satisfied simultaneously:
\begin{eqnarray}
\biggl(\frac{\partial \beta}{\partial r_{h}}\biggr)_{q=q_c,r_{h}=r_{h}^c}=0, \ \ \ \biggl(\frac{\partial^2 \beta}{\partial r_{h}^2}\biggr)_{q=q_c,r_{h}=r_{h}^c}=0
\label{criticalqc}
\end{eqnarray}
By solving eq. (\ref{criticalqc}), we can find $q_c$ and critical horizon radius $r_{h}^c$. However, due to complicated nature of our black hole geometry, we find it hard to 
get the analytical result. Nevertheless, it is straightforward to solve them numerically. In Table (\ref{tableqc}), we have shown numerical results for $q_c$ for different values of $\gamma$. We find 
that higher value of $\gamma$ reduces the magnitude of $q_c$. It shows that higher value of $\gamma$ makes the line of first order phase transition between black holes
of different sizes shorter. Of course in the $\gamma\rightarrow0$ limit, we get back $q_c$ for RN-AdS black hole.

\begin{table}[ht]
\begin{center}
 \begin{tabular}{|p{2.1cm}|p{2.1cm}|p{2.1cm}|p{2.1cm}|p{2.1cm}|}
\multicolumn{5}{c}
 {Numerical results for $q_c$ at order $\gamma$}\\
\hline
$\gamma$ &D=4&D=5&D=6&D=7 \\
\hline
$0$ &0.1667&0.0861&0.0571&0.0424 \\
\hline
$0.0001$ &0.1663&0.0858&0.0568&0.0421 \\
\hline
$0.001$ &0.1628&0.0835&0.0546&0.0399 \\
\hline
$0.002$ &0.1592&0.0811&0.0524&0.0377 \\
\hline
$0.003$ &0.1559&0.0789&0.0505&0.0359 \\
\hline
$0.004$ &0.1528&0.0769&0.0487&0.0343 \\
\hline
$0.005$ &0.1498&0.0750&0.0472&0.0329 \\
\hline
\multicolumn{5}{c}{ Numerical results for $q_c$ at order $\gamma^2$} \\
\hline
$\gamma$ &D=4 &D=5&D=6 &D=7 \\
\hline
$0$ &0.1667&0.0861&0.0571&0.0424 \\
\hline
$0.0001$ &0.1663&0.0858&0.0568&0.0421 \\
\hline
$0.001$ &0.1626&0.0833&0.0543&0.0395 \\
\hline
$0.002$ &0.1585&0.0806&0.0512&0.0363 \\
\hline
$0.003$ &0.1544&0.0778&0.0485&0.0333 \\
\hline
$0.004$ &0.1503&0.0749&0.0455&0.0306 \\
\hline
$0.005$ &0.1463&0.0721&0.0428&0.0279 \\
\hline
\end{tabular}
\caption{
Numerical results for $q_c$ for different
values of the Weyl coupling $\gamma$ and spacetime dimension $D$. Top and bottom part of the table show results for $q_c$ at $\gamma$ and $\gamma^2 $ order respectively.} 
\label{tableqc}
\end{center}
\end{table}

In order to justify and further validate our result for $q_c$ at linear order in $\gamma$, it will be useful if we can find change in $q_c$ when the next to leading order 
corrections due to $\gamma$ are taken into account. In table (\ref{tableqc}), we have provided results for $q_c$ at $\gamma^2$ order. We see that deviation of $q_c$ from leading order is 
at third decimal or, at most, is at second decimal place. These small corrections at $\gamma^2$ order further justify and substantiate our perturbative analysis at the leading order. 
In the next section,  we will also show that even the QNMs of scalar field perturbation do not change significantly when $\gamma^2$ order corrections are taken into account.  
%%%%%%%%%%%%%%%%%%%%%%%%%%%%%%
\begin{figure}[t!]
\begin{minipage}[b]{0.5\linewidth}
\centering
\includegraphics[width=2.8in,height=2.3in]{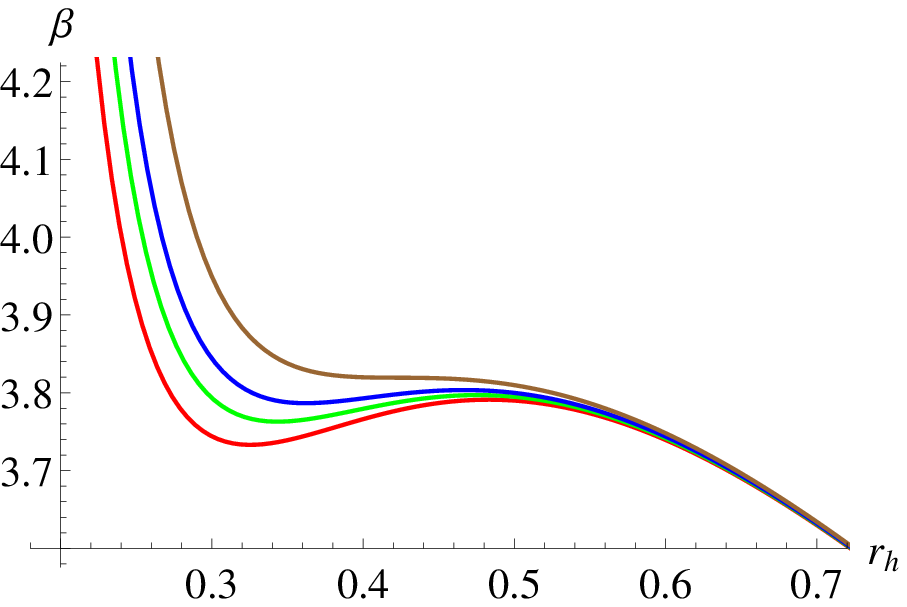}
\caption{ \small $\beta$ as a function of horizon radius for $q=0.15$ in four dimensions. Red, green, blue and brown curves correspond to 
$\gamma$ = $0.001$, $0.002$, $0.003$ and $0.005$ respectively.}
\label{4DbetaVsRhqpt15Vsgamma}
\end{minipage}
\hspace{0.4cm}
\begin{minipage}[b]{0.5\linewidth}
\centering
\includegraphics[width=2.8in,height=2.3in]{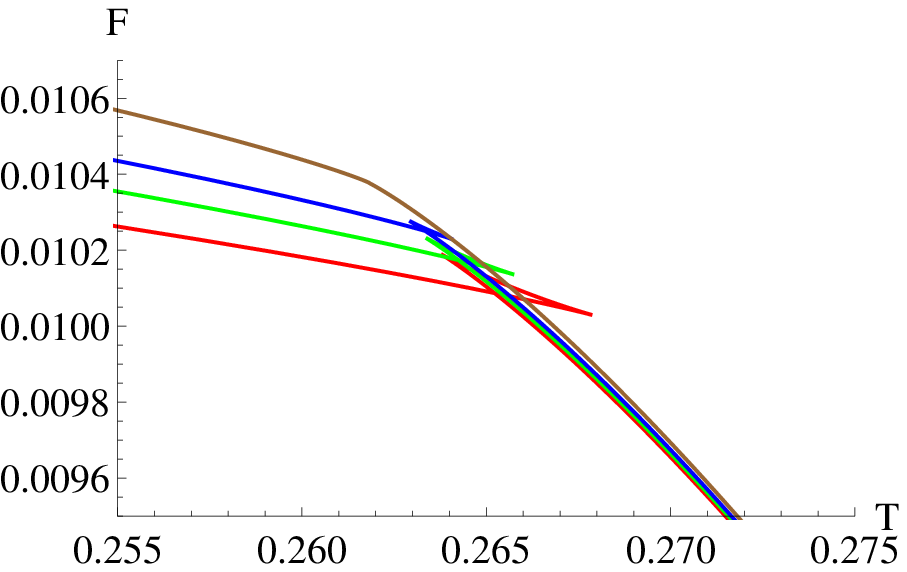}
\caption{\small Free energy as a function of temperature for $q=0.15$ in four dimensions.  Red, green, blue and brown curves correspond to 
$\gamma$ = $0.001$, $0.002$, $0.003$ and $0.005$ respectively.}
\label{4DFVsTqpt15Vsgamma}
\end{minipage}
\end{figure}
%%%%%%%%%%%%%%%%%%%%%%%%%%%%%%

To complete our analysis for four dimensions, in figs.(\ref{4DbetaVsRhqpt15Vsgamma}) and (\ref{4DFVsTqpt15Vsgamma}), we have shown $\beta$ vs $r_h$ and $F$ vs $T$ plots for fixed $q=0.15$. 
Here red, green, blue and cyan curves 
correspond to $\gamma$=$0.001$, $0.002$, $0.003$ and $0.005$ respectively. For small value of $\gamma$, say $\gamma=0.002$, we again find first order like phase transition which
disappears at higher values. This is again consistent with our earlier findings and also suggests that for a fixed $q$, $\gamma$ can equivalently
control the nature of the black hole phase transition.
\\
\\
%%%%%%%%%%%%%%%%%%%%%%%%%%%%%%
\begin{figure}[t!]
\begin{minipage}[b]{0.5\linewidth}
\centering
\includegraphics[width=2.8in,height=2.3in]{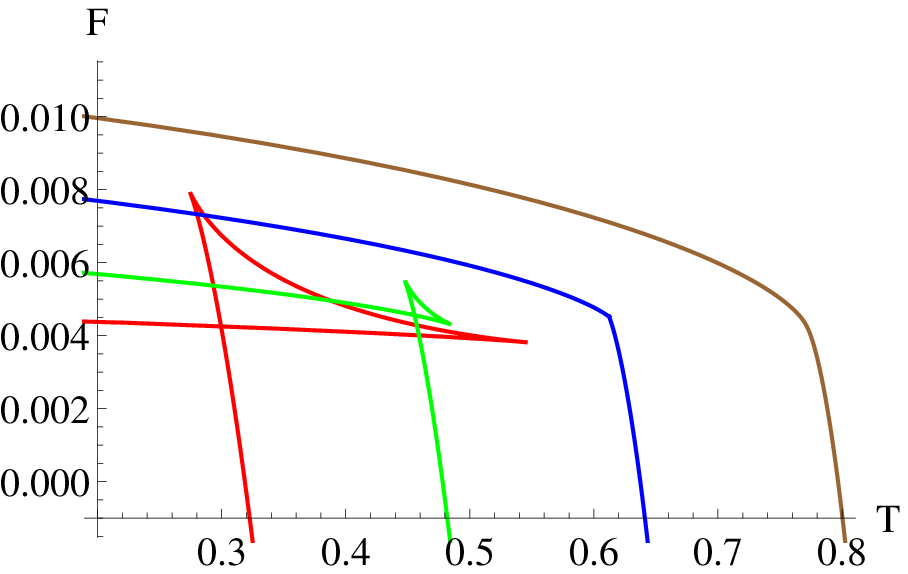}
\caption{ \small Free energy as a function of temperature for  $q=1/20$ and  $\gamma=0.001$ in various dimensions.  Red, green, blue and brown curves correspond to 
$D$ = $4$, $5$, $6$ and $7$ respectively.}
\label{FVsTgamma1by1000q1by20VsD}
\end{minipage}
\hspace{0.4cm}
\begin{minipage}[b]{0.5\linewidth}
\centering
\includegraphics[width=2.8in,height=2.3in]{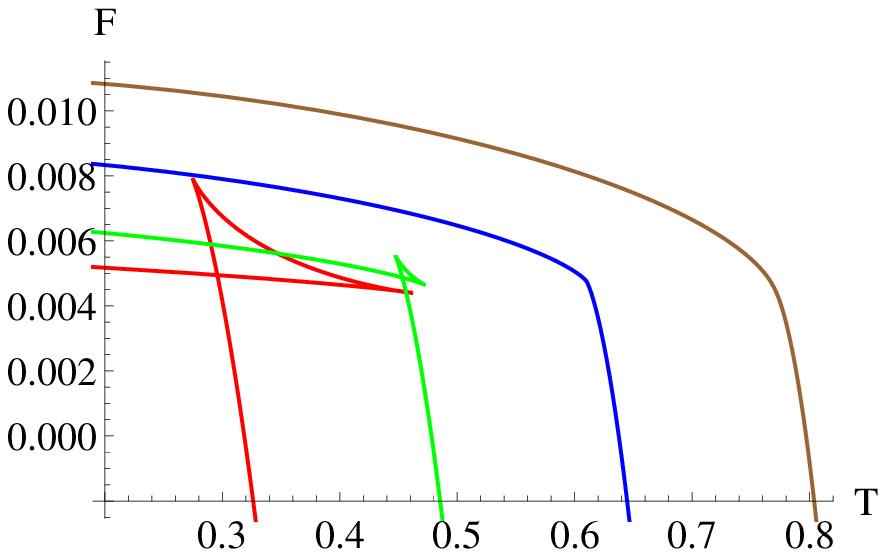}
\caption{\small Free energy as a function of temperature for  $q=1/20$ and  $\gamma=0.004$ in various dimensions.  Red, green, blue and brown curves correspond to 
$D$ = $4$, $5$, $6$ and $7$ respectively.}
\label{FVsTgamma4by1000q1by20VsD}
\end{minipage}
\end{figure}
%%%%%%%%%%%%%%%%%%%%%%%%%%%%%%
Having thoroughly studied the phase structure of Weyl corrected black hole geometry in four dimensions, we now proceed to discuss its properties in higher dimensions. This is shown 
in figs.(\ref{FVsTgamma1by1000q1by20VsD}) and (\ref{FVsTgamma4by1000q1by20VsD}), where free energy as a function of temperature, for two different values of $\gamma$, is shown for 
higher spacetime dimensions. Here we choose $q=1/20$ with red, green, blue and brown curves correspond to $D$=$4$, $5$, $6$ and $7$ respectively.

We find that the overall phase structure of Weyl corrected black hole in higher dimensions is identical to that of the four dimension. We again find $q_c$ below which first order
phase transition from small to large black hole takes place as we increase the temperature. However, the magnitude of $q_c$ decreases as we increase the number of spacetime dimensions.
This can be see from fig.(\ref{FVsTgamma1by1000q1by20VsD}) where for $D=7$ we do not find swallow tail structure in free energy, as opposed to $D<7$. As in four dimensions, here again,
$q_c$ decreases with $\gamma$. An overall dependence of $q_c$ on the number of spacetime dimensions is shown in table (\ref{tableqc}). Moreover, in table (\ref{tableqc}), we have also presented 
results with $\gamma^2$ corrections. We again find only small change in the magnitude of $q_c$ when $\gamma^2$ order corrections are considered.

\section{Scalar perturbation, QNMs and phase transition}

In this section, we will study the dynamics of massless scalar field perturbation in the background geometry of Weyl corrected black hole in various dimensions. We will work in the probe 
limit, where backreaction of scalar field on the background geometry will be neglected. The main aim of this section is to see whether the signature of thermodynamical phase 
transition of Weyl corrected black hole, explored in the previous section, can be reflected in the quasinormal modes of massless scalar field.

We start with the Klein-Gordon equation 
$$\partial_{\mu}\biggl[\sqrt{-g}\partial^{\mu}\Psi \biggr]=0$$
writing $\Psi=\psi(r)e^{-i \omega t} Y[\theta,\phi]$ and using eq.(\ref{metric ansatz}), we get
\begin{eqnarray}
\psi''(r)+\psi'(r)\biggl[{D-2 \over r}+ {f'(r)\over f(r)} -\chi'(r)\biggr]+ \biggl({\omega^2 e^{2\chi}\over f(r)^2}-{l(l+D-3)\over r^2 f(r)}\biggr)\psi(r) =0
\label{psieom}
\end{eqnarray}
We need to solve eq.(\ref{psieom}) with proper boundary conditions. One natural choice is to impose ingoing wave boundary condition at the horizon. This is mathematically 
equivalent to requiring $\psi\propto(r-r_h)^{-i\omega/4 \pi T}$. In order to impose this boundary condition, we use
$$\psi(r)=\psi(r) e^{-i \omega \int {dr\over f(r) e^{-\chi(r)}} }$$
where, one can easily show that $e^{-i \omega \int {dr\over f(r) e^{-\chi(r)}} }$ approaches $(r-r_h)^{-i\omega/4 \pi T}$ near the horizon. We can now rewrite eq.(\ref{psieom}) as
\begin{eqnarray}
\psi''(r)+\psi'(r)\biggl[{D-2 \over r}+ {f'(r)\over f(r)}-{2 i \omega e^{\chi} \over f(r)} - \chi'(r)\biggr] 
-\biggl({(D-2)i \omega e^{\chi} \over r f(r)} \nonumber\\  + {l(l+D-3)\over r^2 f(r)}  \biggr) \psi(r)=0
\label{psieom1}
\end{eqnarray}
We impose second boundary condition at the asymptotic boundary $r\rightarrow\infty$, where we demand $\psi(r\rightarrow\infty)=0$. With these boundary conditions, we numerically solve
eq.(\ref{psieom1}) and find frequencies of the quasinormal modes. For numerical calculations we use the shooting method \footnote{A Mathematica code is available upon request.}.
%%%%%%%%%%%%%%%%%%%%%%%%%%%%%%
\begin{figure}[t!]
\begin{minipage}[b]{0.5\linewidth}
\centering
\includegraphics[width=2.8in,height=2.3in]{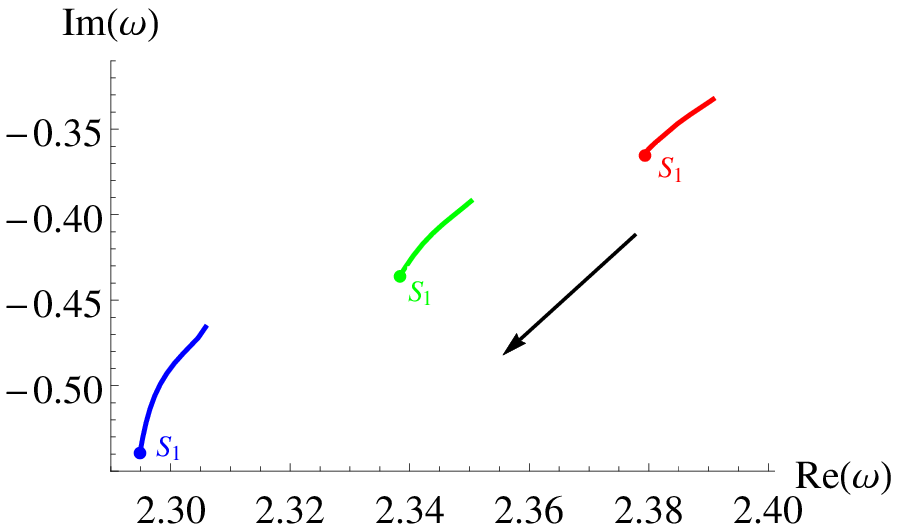}
\caption{ \small QNMs in the small black hole phase. Here $\gamma=0.001$ and red, green and blue curves correspond to $q$=$1/10$, $1/9$ and $1/8$ respectively.
The black arrow indicates the direction of increase of black hole horizon radius.}
\label{SmallBHImwVsRewgamma1by1000vsq}
\end{minipage}
\hspace{0.4cm}
\begin{minipage}[b]{0.5\linewidth}
\centering
\includegraphics[width=2.8in,height=2.3in]{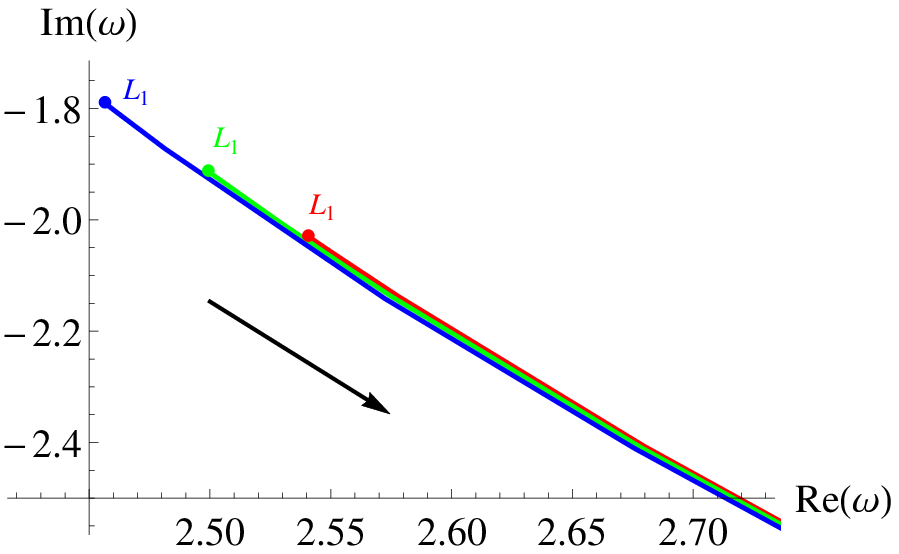}
\caption{\small QNMs in the large black hole phase. here $\gamma=0.001$ and red, green and blue curves correspond to $q$=$1/10$, $1/9$ and $1/8$ respectively.
The black arrow indicates the direction of increase of black hole horizon radius.}
\label{LargeBHImwVsRewgamma1by1000vsq}
\end{minipage}
\end{figure}
%%%%%%%%%%%%%%%%%%%%%%%%%%%%%% 
%%%%%%%%%%%%%%%%%%%%%%%%%%%%
\begin{table}
\centering
\makebox[0pt][c]{\parbox{1.2\textwidth}{%
\begin{minipage}[b]{0.265\hsize}\centering
\begin{tabular}{ | c | c |c | }
\multicolumn{3}{c}
{$q=1/10$} \\
\hline
T & $r_h$ & $\omega$ \\ 
\hline
0.233 & 0.135 & 2.38486-0.34667I  \\ 
0.256 & 0.140 & 2.38092-0.35837I \\
0.273 & 0.145 & 2.38049-0.35982I \\ 
0.278 & 0.146 & 2.38007-0.36130I \\ 
0.281 & 0.147 & 2.37964-0.36282I \\ 
\hline
0.284 & 0.761 & 2.54150-2.03209I \\
0.285 & 0.765 & 2.54515-2.04292I \\
0.289 & 0.800 & 2.57799-2.13761I \\
0.302 & 0.900 & 2.67987-2.40712I \\
0.312 & 1.000 & 2.79214-2.67552I \\
\hline
\end{tabular}
\end{minipage}
\hfill
\begin{minipage}[b]{0.265\hsize}\centering
\begin{tabular}{ | c | c | c | }
\multicolumn{3}{c}
{$q=1/9$} \\
\hline
T & $r_h$ & $\omega$ \\
\hline
0.227 & 0.150 & 2.34385-0.41094I  \\ 
0.245 & 0.155 & 2.34215-0.41723I \\
0.260 & 0.160 & 2.34056-0.42416I \\ 
0.271 & 0.165 & 2.33908-0.43175I \\ 
0.276 & 0.167 & 2.33852-0.43497I \\ 
\hline
0.280 & 0.720 & 2.50248-1.92299I \\
0.283 & 0.750 & 2.52893-2.00424I \\
0.286 & 0.780 & 2.55660-2.08533I \\
0.289 & 0.800 & 2.57568-2.13931I \\
0.302 & 0.900 & 2.67807-2.40842I \\
\hline
\end{tabular}
\end{minipage}
\hfill
\begin{minipage}[b]{0.265\hsize}\centering
\begin{tabular}{ | c | c | c |}
\multicolumn{3}{c}
{$q=1/8$} \\
\hline
T & $r_h$ & $\omega$ \\ 
\hline
0.246 & 0.180 & 2.29733-0.50611I  \\ 
0.255 & 0.185 & 2.29654-0.51364I \\
0.263 & 0.190 & 2.29589-0.52172I \\ 
0.269 & 0.195 & 2.29534-0.53033I \\ 
0.273 & 0.199 & 2.29496-0.53758I \\ 
\hline
0.275 & 0.675 & 2.46067-1.80426I \\
0.277 & 0.700 & 2.48131-1.87194I \\
0.288 & 0.800 & 2.57245-2.14170I \\
0.302 & 0.900 & 2.67555-2.41027I \\
0.317 & 1.000 & 2.78870-2.67796I \\
\hline
\end{tabular}
\end{minipage}%
}}
\caption{Numerical results for quasinormal modes in $D=4$ for $\gamma=0.001$. The left part of the table corresponds to $q=1/10$, middle part corresponds to $q=1/9$ and the right part corresponds $q=1/8$. In all cases
the lower part, below the horizontal line, is for the large black hole phase, while the
upper part is for the small black hole phase.} 
\label{4DQNMgammapPt1001vsq}
\end{table}
%%%%%%%%%%%%%%%%%%%%%%%%%%%%%%%%%%%
\\
\\
In fig.(\ref{SmallBHImwVsRewgamma1by1000vsq}), we have shown the variation of real and imaginary part of the quasinormal frequencies, 
with $\gamma=0.001$ and $l=0$,  for different temperatures in the small black hole phase. Here red, green and blue curves correspond to $q$= $1/10$, $1/9$ and $1/8$ respectively. The black arrow
indicates the direction of increase of black hole temperature (or $r_h$) and points $S_1$ indicate temperature just below the critical temperature $T_*$. We see that with increase in temperature 
the $Re(\omega)$ decreases, whereas the absolute value
of $Im(\omega)$ increases. This implies that slope in the $Re(\omega)$-$Im(\omega)$ plane is negative. 

However, the nature of quasinormal modes changes completely in the large black hole phase. This is shown in fig.(\ref{LargeBHImwVsRewgamma1by1000vsq}), where slope in the
$Re(\omega)$-$Im(\omega)$ plane is now positive. Here, same color coding as in fig.(\ref{SmallBHImwVsRewgamma1by1000vsq}) have been used and points $L_1$ indicate temperature
just above $T_*$. We see that here, as contrast to small black hole 
phase,
with increase in temperature both $Re(\omega)$ and absolute value of $Im(\omega)$ increases. It shows that dynamical perturbation of massless scalar field around black hole horizon
does behave differently in different black hole phases, and therefore, corresponding quasinormal frequencies can be a good measure to probe different phases of the black hole. 

In Table (\ref{4DQNMgammapPt1001vsq}), we have shown numerical values of the quasinormal frequencies. The left part, 
the middle part and right part of the table correspond to $q=1/10$, $q=1/9$ and  $q=1/8$ respectively. In all cases,
the lower part below the horizontal line is for large black hole phase, while the upper part is for small black hole phase. We find that as we decrease $r_h$ in small 
black hole phase the $Re(\omega)$ increases whereas the absolute value of $Im(\omega)$ decreases \footnote{The decrease in $Im(\omega)$ as we decrease $r_{h}$
is expected on the general ground, since the absorption of the scalar field is caused by 
presence of the black hole horizon. Therefore as we decrease horizon radius we expect less absorption and hence less $Im(\omega)$ \cite{Liu}.}. 
For large black hole phase temperature increases with $r_{h}$. In this phase both $Re(\omega)$ and absolute value
of $Im(\omega)$ increases with $r_{h}$. One can clearly observe change in 
the pattern of $\omega$ near the critical temperature.
%%%%%%%%%%%%%%%%%%%%%%%%%%%%%%
\begin{figure}[h!]
\begin{minipage}[b]{0.5\linewidth}
\centering
\includegraphics[width=2.8in,height=2.3in]{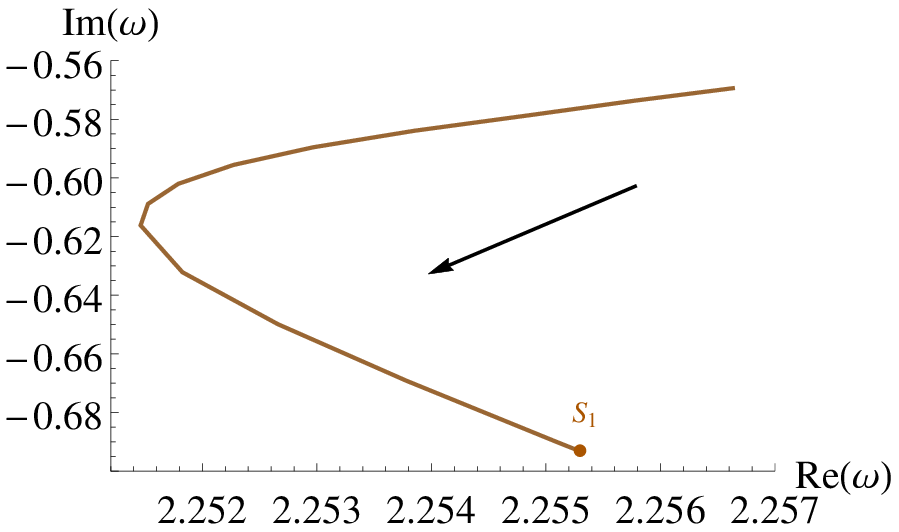}
\caption{ \small QNMs for small black hole with $\gamma=0.001$ and $q$=$1/7$.
The black arrow indicates the direction of increase of black hole horizon radius. The brown dot indicates temperature just below the critical temperature.}
\label{SmallBHImwVsRewgamma1by1000q1by7}
\end{minipage}
\hspace{0.4cm}
\begin{minipage}[b]{0.5\linewidth}
\centering
\includegraphics[width=2.8in,height=2.3in]{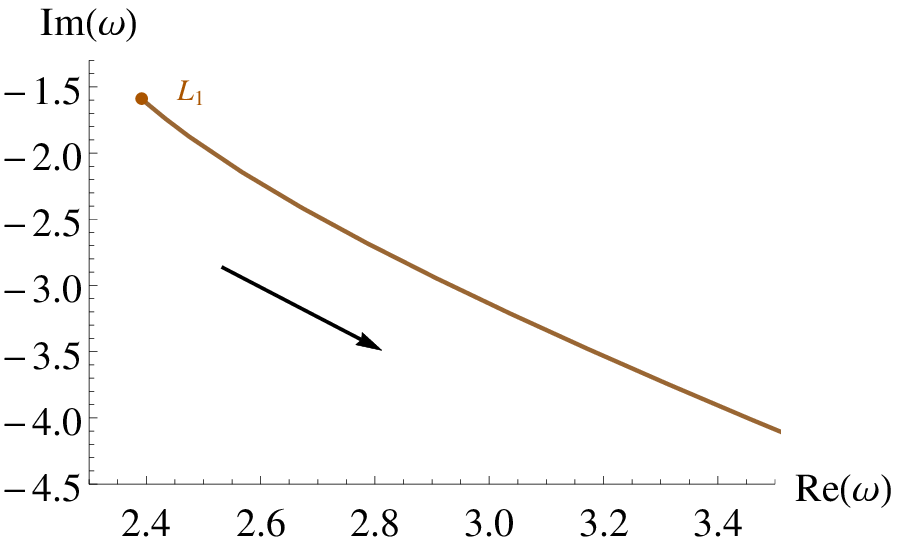}
\caption{\small QNMs for large black hole with $\gamma=0.001$ and $q$=$1/7$.
The black arrow indicates the direction of increase of black hole horizon radius. The brown dot indicates temperature just above the critical temperature.}
\label{LargeBHImwVsRewgamma1by1000q1by7}
\end{minipage}
\end{figure}
%%%%%%%%%%%%%%%%%%%%%%%%%%%%%% 

The contrasting nature of the quasinormal modes in small and large black hole phases does illustrate its effectiveness 
to probe the black hole phase transition. However, there is one subtle point which needs to be mentioned. We find that as $q$ approaches the second order critical 
point $q_c$, quasinormal modes
are not as much effective to probe the black hole phase transition as in the case when $q\ll q_c$. This is shown in figs.(\ref{SmallBHImwVsRewgamma1by1000q1by7}) and 
(\ref{LargeBHImwVsRewgamma1by1000q1by7}),
where we clearly see that for $q=1/7$,
$Re(\omega)$-$Im(\omega)$ plane does not show different slope in small and large black hole phases near the critical temperature. The quasinormal modes in large black hole phase continues to have same
characteristic feature as in the previous case, however now, in small black hole phase they have positive slope near the critical temperature. Although, we do get negative slope (or change
in slope) in small black hole phase but it appears well below the critical temperature. The same story exists for $\gamma=0$ case too, which corresponds to RN-AdS black hole, and 
therefore are not special to Weyl corrected black holes \footnote{In there analysis, \cite{Liu} did not mention these subtle points.}.
%%%%%%%%%%%%%%%%%%%%%%%%%%%%%%
\begin{figure}[t!]
\begin{minipage}[b]{0.5\linewidth}
\centering
\includegraphics[width=2.8in,height=2.3in]{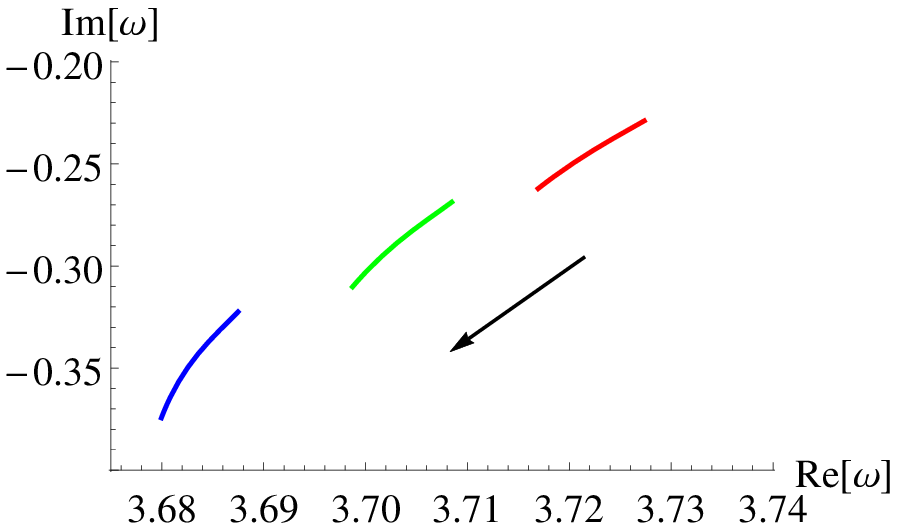}
\caption{ \small QNMs for small black hole in $D=5$ with $\gamma=0.002$. Red, green and blue curves correspond to $q$=$1/40$, $1/35$ and $1/30$ respectively.
The black arrow indicates the direction of increase of black hole horizon radius.}
\label{5DSmallBHImwVsRewgamma1by500Vsq}
\end{minipage}
\hspace{0.4cm}
\begin{minipage}[b]{0.5\linewidth}
\centering
\includegraphics[width=2.8in,height=2.3in]{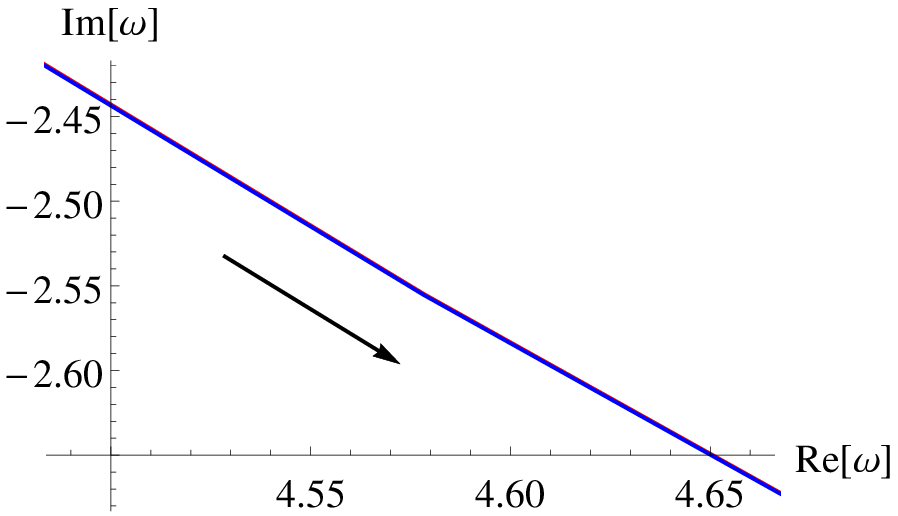}
\caption{\small QNMs for large black hole in $D=5$ with $\gamma=0.002$. Red, green and blue curves correspond to $q$=$1/40$, $1/35$ and $1/30$ respectively.
The black arrow indicates the direction of increase of black hole horizon radius.}
\label{5DLargeBHImwVsRewgamma1by500Vsq}
\end{minipage}
\end{figure}
%%%%%%%%%%%%%%%%%%%%%%%%%%%%%%   
%%%%%%%%%%%%%%%%%%%%%%%%%%%%%%
\begin{figure}[t!]
\begin{minipage}[b]{0.5\linewidth}
\centering
\includegraphics[width=2.8in,height=2.3in]{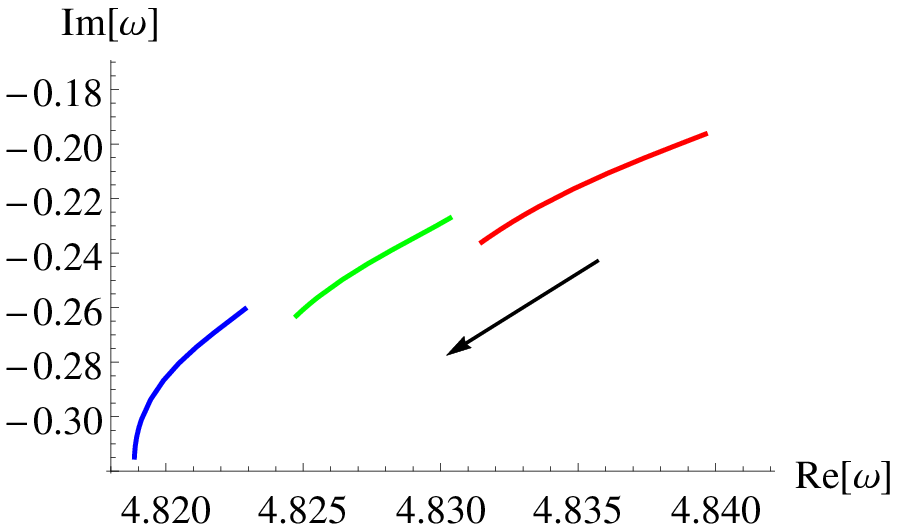}
\caption{ \small QNMs for small black hole in $D=6$ with $\gamma=0.002$. Red, green and blue curves correspond to $q$=$1/100$, $1/85$ and $1/75$ respectively.
The black arrow indicates the direction of increase of black hole horizon radius.}
\label{6DSmallBHImwVsRewgamma1by500Vsq}
\end{minipage}
\hspace{0.4cm}
\begin{minipage}[b]{0.5\linewidth}
\centering
\includegraphics[width=2.8in,height=2.3in]{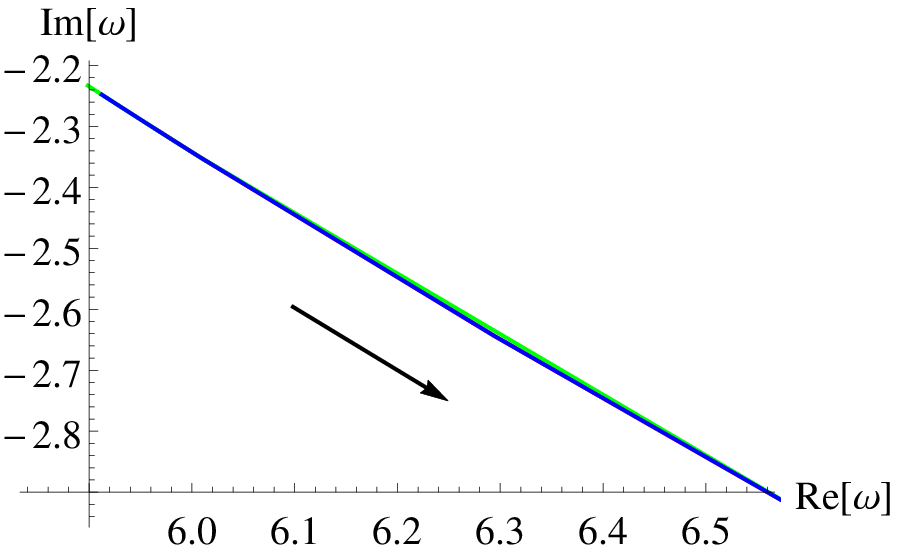}
\caption{\small QNMs for large black hole in $D=6$ with $\gamma=0.002$. Red, green and blue curves correspond to $q$=$1/100$, $1/85$ and $1/75$ respectively.
The black arrow indicates the direction of increase of black hole horizon radius.}
\label{6DLargeBHImwVsRewgamma1by500Vsq}
\end{minipage}
\end{figure}
%%%%%%%%%%%%%%%%%%%%%%%%%%%%%%   
%%%%%%%%%%%%%%%%%%%%%%%%%%%%%% 

In the light of above discussion an important question which naturally arises is to find a condition which allows quasinormal modes to effectively probe the phase transition. Currently we don't have 
any concrete
answer for this condition. However, from our preliminary analysis, it appears to us that the condition should be related to separation between $r_{h_{L_1}}$ and $r_{h_{S_1}}$. 
For example, we see that as $q$ approaches $q_c$ the distance $r_{h_{L_1}}-r_{h_{S_1}}$ decreases and becomes zero at $q_c$. In order for quasinormal modes to 
effectively probe the phase transition one would physically expect that
some conditions like $r_{h_{L_1}}-r_{h_{S_1}}>Abs[\omega]$ should satisfy. Numerically, we checked that in our system if the condition
\begin{equation}
(r_{h_{L_1}}-r_{h_{S_1}}) - \biggl(\frac{Abs[\omega\rvert_{r_{h_{S_1}}}]r_{h_{S_1}}}{Abs[\omega\rvert_{r_{h_{L_1}}}]r_{h_{L_1}}} \biggr) \gtrsim 0.3
\label{QNMcondition}
\end{equation}
is satisfied then quasinormal modes do show different slope in small and large black hole phases 
\footnote{We introduce $r_h$ in the second term of eq.(\ref{QNMcondition}) in order to make it dimensionless. For fixed $q$ and $L$, the only remaining scale in the system is 
horizon radius $r_h$.}. 
Here, $\omega\rvert_{r_{h_{S_1}}}(\omega\rvert_{r_{h_{L_1}}})$ are $\omega$ values evaluated at
$r_h={r_{h_{S_1}}}({r_{h_{L_1}}})$. We have checked the criteria of eq.(\ref{QNMcondition}) for several values of $\gamma$ and $q$ (provided it is less than $q_c$) and found consistent 
results. However, apart from this numerical analysis we do not have any substantial justification of eq.(\ref{QNMcondition}) and therefore more detailed and
careful study of quasinormal modes near the second order critical point is required.

\begin{table}
\centering
\makebox[0pt][c]{\parbox{1.2\textwidth}{%
\begin{minipage}[b]{0.265\hsize}\centering
\begin{tabular}{ | c | c |c |}
\multicolumn{3}{c}
{$q=1/40$} \\
\hline
T & $r_h$ & $\omega$ \\ 
\hline
0.297 & 0.190 & 3.72421-0.23785I  \\ 
0.437 & 0.205 & 3.71849-0.25624I \\
0.443 & 0.206 & 3.71811-0.25766I \\ 
0.450 & 0.207 & 3.71772-0.25911I \\ 
0.456 & 0.208 & 3.71733-0.26058I \\ 
0.462 & 0.209 & 3.71695-0.26207I \\ 
\hline
0.466 & 0.920 & 4.41734-2.32192I \\
0.467 & 0.930 & 4.43698-2.35116I \\
0.468 & 0.935 & 4.44686-2.36577I \\
0.469 & 0.945 & 4.46673-2.39498I \\
0.477 & 1.000 & 4.57857-2.55515I \\
0.716 & 2.000 & 7.06844-5.39070I \\
\hline
\end{tabular}
\end{minipage}
\hfill
\begin{minipage}[b]{0.265\hsize}\centering
\begin{tabular}{ | c | c | c | }
\multicolumn{3}{c}
{$q=1/35$} \\
\hline
T & $r_h$ & $\omega$ \\
\hline
0.145 & 0.190 & 3.70846-0.26873I  \\ 
0.370 & 0.210 & 3.70308-0.28874I \\
0.436 & 0.220 & 3.70013-0.30246I \\ 
0.452 & 0.223 & 3.69928-0.30702I \\ 
0.457 & 0.224 & 3.69900-0.30859I \\ 
0.462 & 0.225 & 3.69873-0.31018I \\ 
\hline
0.465 & 0.915 & 4.40749-2.30749I \\
0.466 & 0.920 & 4.41725-2.32212I \\
0.467 & 0.930 & 4.43690-2.35135I \\
0.470 & 0.950 & 4.47664-2.40975I \\
0.477 & 1.000 & 4.57850-2.55530I \\
0.716 & 2.000 & 7.06844-5.39071I \\
\hline
\end{tabular}
\end{minipage}
\hfill
\begin{minipage}[b]{0.265\hsize}\centering
\begin{tabular}{ | c | c | c |}
\multicolumn{3}{c}
{$q=1/30$} \\
\hline
T & $r_h$ & $\omega$ \\ 
\hline
0.228 & 0.210 & 3.68666-0.32664I  \\ 
0.321 & 0.220 & 3.68462-0.33698I \\
0.438 & 0.240 & 3.68082-0.36447I \\ 
0.453 & 0.244 & 3.68023-0.37100I \\ 
0.457 & 0.245 & 3.68100-0.37268I \\ 
0.460 & 0.246 & 3.67996-0.37438I \\ 
\hline
0.463 & 0.900 & 4.37828-2.26390I \\
0.464 & 0.905 & 4.38793-2.27854I \\
0.465 & 0.915 & 4.40735-2.30780I \\
0.470 & 0.950 & 4.47652-2.41002I \\
0.477 & 1.000 & 4.57841-2.55552I \\
0.716 & 2.000 & 7.06843-5.39072I \\
\hline
\end{tabular}
\end{minipage}%
}}
\caption{
Numerical results for quasinormal modes in $D=5$ for $\gamma=0.002$. The left part of the table corresponds to $q=1/40$, middle part corresponds to $q=1/35$ and the right part corresponds $q=1/30$. 
In all cases the lower part, below the horizontal line, is for the large black hole phase, while the
upper part is for the small black hole phase.} 
\label{5DQNMgammaPt002vsq}
\end{table}
%%%%%%%%%%%%%%%%%%%%%%%%%%%%%%%
%%%%%%%%%%%%%%%%%%%%%%%%%%%%%% 
\begin{table}
\centering
\makebox[0pt][c]{\parbox{1.2\textwidth}{%
\begin{minipage}[b]{0.265\hsize}\centering
\begin{tabular}{ | c | c |c |}
\multicolumn{3}{c}
{$q=1/100$} \\
\hline
T & $r_h$ & $\omega$ \\ 
\hline
0.275 & 0.240 & 4.83858-0.20046I  \\ 
0.437 & 0.250 & 4.83610-0.21062I \\
0.552 & 0.260 & 4.83356-0.22303I \\ 
0.604 & 0.266 & 4.83215-0.23148I \\ 
0.619 & 0.268 & 4.83171-0.23446I \\ 
0.627 & 0.269 & 4.83150-0.23598I \\ 
\hline
0.633 & 0.975 & 5.93952-2.27731I \\
0.634 & 0.985 & 5.96694-2.30681I \\
0.635 & 0.990 & 5.98072-2.32156I \\
0.637 & 1.000 & 6.00841-2.35101I \\
0.756 & 1.500 & 7.56677-3.79247I \\
0.915 & 2.000 & 9.33354-5.19484I \\
\hline
\end{tabular}
\end{minipage}
\hfill
\begin{minipage}[b]{0.265\hsize}\centering
\begin{tabular}{ | c | c | c |}
\multicolumn{3}{c}
{$q=1/85$} \\
\hline
T & $r_h$ & $\omega$ \\
\hline
0.397 & 0.260 & 4.82734-0.24396I  \\ 
0.510 & 0.270 & 4.82555-0.25612I \\
0.592 & 0.280 & 4.82407-0.27035I \\ 
0.611 & 0.283 & 4.82371-0.27501I \\ 
0.617 & 0.284 & 4.82361-0.27660I \\ 
0.623 & 0.285 & 4.82350-0.27821I \\ 
\hline
0.631 & 0.960 & 5.89873-2.23303I \\
0.632 & 0.970 & 5.92587-2.26258I \\
0.633 & 0.980 & 5.95320-2.29210I \\
0.637 & 1.000 & 6.00841-2.35104I \\
0.756 & 1.500 & 7.56677-3.79247I \\
0.915 & 2.000 & 9.33354-5.19484I \\
\hline
\end{tabular}
\end{minipage}
\hfill
\begin{minipage}[b]{0.265\hsize}\centering
\begin{tabular}{ | c | c | c |}
\multicolumn{3}{c}
{$q=1/75$} \\
\hline
T & $r_h$ & $\omega$ \\ 
\hline
0.392 & 0.270 & 4.82111-0.27500I  \\ 
0.497 & 0.280 & 4.81991-0.28680I \\
0.573 & 0.290 & 4.81911-0.30114I \\ 
0.598 & 0.294 & 4.81894-0.30743I \\ 
0.609 & 0.296 & 4.81889-0.31069I \\ 
0.619 & 0.298 & 4.81886-0.31402I \\ 
\hline
0.631 & 0.965 & 5.91228-2.24785I \\
0.632 & 0.970 & 5.92587-2.26262I \\
0.633 & 0.985 & 5.95320-2.29213I \\
0.637 & 1.000 & 6.00841-2.35107I \\
0.760 & 1.500 & 7.56677-3.79248I \\
0.915 & 2.000 & 9.33354-5.19484I \\
\hline
\end{tabular}
\end{minipage}%
}}
\caption{
Numerical results for quasinormal modes in $D=6$ for $\gamma=0.002$. The left part of the table corresponds to $q=1/100$, middle part corresponds to $q=1/85$ and the right part corresponds $q=1/75$. 
In all cases the lower part, below the horizontal line, is for the large black hole phase, while the
upper part is for the small black hole phase.} 
\label{6DQNMgammaPt002vsq}
\end{table}
%%%%%%%%%%%%%%%%%%%%%%%%%%%%%%%%%%%%%%%%%%%%%

Now, we move on to discuss quasinormal modes for higher dimensional black holes. The results are shown in figs. (\ref{5DSmallBHImwVsRewgamma1by500Vsq})-(\ref{6DLargeBHImwVsRewgamma1by500Vsq}).
The overall
characteristic features of the quasinormal modes are found to be same as in the four dimensions, and therefore we will be very brief here. We again found
change in the slope of quasinormal modes (as long as we are away from the critical charge) as we decrease the temperature below the critical temperature. The numerical values of 
the quasinormal modes are given in 
Table (\ref{5DQNMgammaPt002vsq}) and (\ref{6DQNMgammaPt002vsq}), where dependence of quasinormal modes on charge and  spacetime dimensions are established. We found
that, for fixed ($q$, $\gamma$ and $r_h$), as we increase the number of spacetime dimensions the $Re(\omega)$ increases whereas the absolute value of 
$Im(\omega)$ decreases. It implies that for fixed $\gamma$ and $q$, the damping time $(\tau \propto 1/Im(\omega))$ increases with dimensions. According to the
gauge/gravity duality, this means that in higher dimensions, it will take more time for the quasinormal ringing to settle down to the thermal equilibrium.

\section{Conclusions}
In this section, we summarize the main results of our work. We have studied thermodynamics and quasinormal modes of scalar field perturbation for $D$ dimensional charged 
black hole in the presence of Weyl coupling. We started with AdS gravity which couples to the gauge field through two and four derivative interaction terms. We treated 
coefficient $\gamma$ of four derivative interaction term as a perturbative parameter and solved Einstein-Maxwell equations order by order in $\gamma$. We first explicitly 
constructed charged black hole solution in $D$ dimensions at linear order in $\gamma$ and further outlined the necessary steps for black hole solution at $\gamma^2$ order (in Appendix A).

We then used the black hole solution to construct various thermodynamic quantities and showed that the first law of black hole thermodynamics is satisfied in all spacetime 
dimensions, not just
at the leading order but also at the subleading order in $\gamma$. Then, we studied black hole thermodynamics in the fixed charge
ensemble and found the analogous Van der Waals liquid-gas type phase transition in it. We found a line of first order phase transition between small and large black hole phases which
terminates at the second order critical point. We  also established the dependence of critical charge on $\gamma$ and showed that $\gamma$ can be a control parameter which 
can tune the order of the phase transition. We then explicitly showed that the our linear order analysis is trustable as nonlinear order corrections are small.

We then reexamined the argument of QNMs could be a probe of the black hole phase transition in our Weyl corrected black hole geometry. We determined the QNMs of massless scalar 
field
perturbation around small and large black hole phases, and found that QNMs changes differently in small and large black hole phases as we increase the horizon radius.  
Our results strengthen the claim made in \cite{Koutsoumbas}-\cite{He} that the QNMs can be used as a dynamical tool to probe the black hole phase transition. However,
we have also highlighted some issues near the second order critical point where QNMs are not as much as effective to probe the phase transition as away from the critical point, and therefore, 
more detailed and careful study of QNMs near the critical point is called for.

We conclude here by pointing out some problems which would be interesting to investigate
in the future. It would be interesting to analyze the QNMs of eletromagnetic and gravitational perturbation in our system. However, since eletromagnetic and gravitational perturbation 
will source other components of the gauge field and metric, their QNMs calculation might be substantially more complicated. We leave this
issue for a future work.

\begin{center}
{\bf Acknowledgements}
\end{center}
I am very grateful to Tapobrata Sarkar for useful discussions and for giving me valuable comments. I would like to thank A. Dey, P. Roy, 
Zodinmawia and A. Sharma for careful reading of the manuscript and pointing out the necessary corrections.

\appendix
\section{Black hole solution with $\gamma^2$ correction}
Here, we present the details of our calculations for correction in metric and gauge field at $\gamma^2$ order. We present the calculations only for four dimensions as the calculations
for higher dimensions can be straightforwardly generalized. We start with the following ansatz for 
$\phi(r)$, $\chi(r)$ and $f(r)$
\begin{eqnarray}
\phi(r) &=& \phi_0(r)+ \gamma \phi_1(r) + \gamma^2 \phi_2(r),  \nonumber\\
\chi(r) &=& \chi_0(r) + \gamma \chi_1(r)+ \gamma^2 \chi_2(r)\, \nonumber\\
f(r)&=&f_0(r)\bigl(1+\gamma \mathcal{F}_{1}(r)+\gamma^2 \mathcal{F}_{2}(r)\bigr)\
\label{perturbationansatz2}
\end{eqnarray}
The forms of zeroth order solution ( $\phi_0(r)$, $\chi_0(r)$ and $f_0(r)$)  and the first order solution ($\phi_1(r)$, $\chi_1(r)$ and $\mathcal{F}_{1}(r)$) are obtained in section 2.
The aim of this Appendix is solve Einstein and Maxwell equations at $\gamma^2$ order and find the forms of second order solutions i.e $\phi_2(r)$, $\chi_2(r)$ and $\mathcal{F}_{2}(r)$. 
The Maxwell and the $(tt)$, $(rr)$ components of Einstein equations at order
$\gamma^2$ give solution for $\chi_2(r)$, $\phi_{2}(r)$ and $\mathcal{F}_{2}(r)$ as
\begin{eqnarray}
\chi_{2}(r) &=& C_2 - {1280 m q^2 L^4 \over 21 r^7} + {1448 L^4 q^4 \over 9 r^8}
\end{eqnarray}
\begin{eqnarray}
\phi_{2}(r) = -{2596 L^4 q^5 \over 45 r^9}+{1696mq^3L^4\over21r^8}-{128qm^2L^4\over7r^7}-{128q^3L^4\over7 r^7} \nonumber\\
-{128L^2 q^3\over5 r^5}-{{C_4}q\over r}
+{4qr_{h}^3 k_1 \over r^4}+C_3
\end{eqnarray}
\begin{eqnarray}
\mathcal{F}_{2}(r) = {1\over f_{0}(r)} \biggl[{10240 L^4 q^6\over 27 r^{10}}-{554 m L^4 q^4\over r^9}+{3328 m^2 q^2 L^4 \over 21 r^8}+{27392 L^4 q^4\over 63 r^8} - \nonumber\\
{512 m q^2 L^4\over3 r^7}+{7936 L^2 q^4\over 15 r^6}-{640 m q^2 L^2\over 3 r^5}-{10q^2 r_{h}^3 k_1\over 3 r^5}+{q^2 C_4\over r^2}+{2 q^2 C_2\over r^2} +{C_1 r_{h}^3 \over L^2 r}
\biggr]
\end{eqnarray}
where $C_1$, $C_2$, $C_3$ and $C_4$ are dimensionless integration constants. These constants can again be determined by imposing similar constrains as in section 2. We omit
the intermediate steps here and simply write the final answer
\begin{eqnarray}
\chi_{2}(r) &=& - {1280 m q^2 L^4 \over 21 r^7} + {1448 L^4 q^4 \over 9 r^8}
\label{chi1EOM2}
\end{eqnarray}
\begin{eqnarray}
 \phi_{2}(r) = -{2596 L^4 q^5 \over 45 r^9}+{1696mq^3L^4\over21r^8}-{128qm^2L^4\over7r^7}-{128q^3L^4\over7 r^7} -{128L^2 q^3\over5 r^5}+{136mqL^2\over 15r^4} \nonumber\\
 +{2596L^2q^5\over 49r_{h}^9}-{1696mq^3L^4\over 21r_{h}^8}+{1976qm^2L^4\over 105r_{h}^7}+{128q^3L^4\over 7r_{h}^7}+{8mqL^4\over 5r_{h}^6}-{32qL^4\over 15r_{h}^5}+{128q^3L^2\over 5r_{h}^5}  \nonumber\\
 -{136mqL^2\over 15r_{h}^4}+{32qL^2\over 5r_{h}^3}-{8qm^2L^4\over 15 r^4r_{h}^3} 
-{8mqL^4\over 5 r^4r_{h}^2}+{128q\over 15r_{h}}+{32qL^4\over 15 r^4r_{h}}-{32L^2qr_{h}\over 5 r^4}-{128qr_{h}^3\over 15r^4}
\label{phi1EOM2}
\end{eqnarray}
\begin{eqnarray}
\mathcal{F}_{2}(r) = {1\over f_{0}(r)} \biggl[{10240 L^4 q^6\over 27 r^{10}}-{554 m L^4 q^4\over r^9}+{3328 m^2 q^2 L^4 \over 21 r^8}+{27392 L^4 q^4\over 63 r^8} - \nonumber\\
{512 m q^2 L^4\over3 r^7}+{7936 L^2 q^4\over 15 r^6}-{1988 m q^2 L^2\over 9 r^5}+{11086m\over 45 r}+{2990m^3L^4\over 189 r r_{h}^6} -{1612 m^2L^4\over 21 r r_{h}^5} \nonumber\\
+{7450mL^4\over 63 r r_{h}^4}+{4 m^2q^2L^4\over 9 r^5 r_{h}^3}-{10832 L^4\over 189 r r_{h}^3}-{37636 m^2L^2\over 315 r r_{h}^3}+{4 mq^2L^4\over 3 r^5 r_{h}^2}
+{38284 mL^2\over 105 r r_{h}^2} \nonumber\\
-{16q^2L^4\over 9 r^5 r_{h}}-{81056 L^2\over 315 r r_{h}}+{16 q^2L^2 r_h\over 3 r^5} -{35984 r_h\over 105 r} + {64 q^2 r_{h}^3\over 9 r^5 }-{19264 r_{h}^3\over 135 L^2 r}     \biggr]
\label{F1EOM2}
\end{eqnarray}
%%%%%%%%%%%%%%%%%%%%%%%%%%%%%%
\begin{figure}[t!]
\begin{minipage}[b]{0.5\linewidth}
\centering
\includegraphics[width=2.8in,height=2.1in]{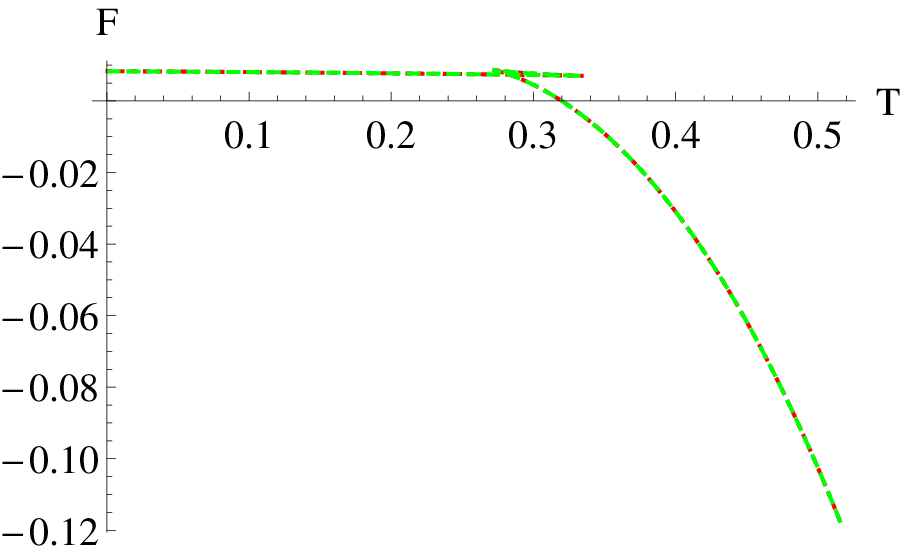}
\caption{ \small Comparison between free energy at order $\gamma$ (dotted red) with free energy at order $\gamma^2$ (dashed green). Here $\gamma=0.001$ and $q=1/10$. }
\label{FVsTgamma1by1000comparison}
\end{minipage}
\hspace{0.4cm}
\begin{minipage}[b]{0.5\linewidth}
\centering
\includegraphics[width=2.8in,height=2.1in]{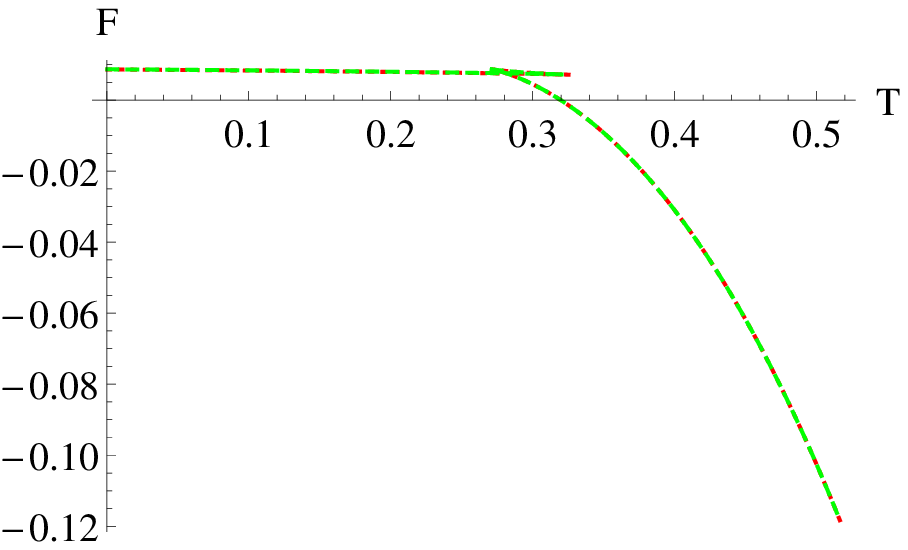}
\caption{\small Comparison between free energy at order $\gamma$ (dotted red) with free energy at order $\gamma^2$ (dashed green). Here $\gamma=0.002$ and $q=1/10$.}
\label{FVsTgamma1by500comparison}
\end{minipage}
\end{figure}
%%%%%%%%%%%%%%%%%%%%%%%%%%%%%%  
%%%%%%%%%%%%%%%%%%%%%%%%%%%%%%  

Using the methodology given in section 3, one can easily compute the black hole mass $M$, Wald entropy $S$ and potential $\Phi$ at $\gamma^2$ order. From these we can calculate
total on-shell action, Gibbs and Helmholtz free energies
\begin{eqnarray}
S_{Total}={\omega_2 \beta \over 16 \pi G_4}\biggl(r_h-r_{h}^3-{q^2\over r_{h}} \biggr)+{\omega_2 \beta q^2 \gamma \over 8 \pi G_4}\biggl({1\over r_h}+{1\over 3 r_{h}^3}-{q^2\over 15 r_{h}^5}  \biggr) \nonumber\\
{\omega_2 \beta q^2 \gamma^2 \over\pi G_4}\biggl({1495 q^4\over 1512 r_{h}^9} -{421 q^2\over 504 r_{h}^7}-{3781 q^2\over 804 r_{h}^5}+{8\over 21 r_{h}^5}
+{80\over 21 r_{h}^3}+{24\over 7 r_{h}} \biggr)
\label{onshellaction2}
\end{eqnarray}
\begin{eqnarray}
G={\omega_2\over 16 \pi G_4}\biggl(r_h-r_{h}^3-{q^2\over r_{h}} \biggr)+{\omega_2 q^2 \gamma \over 8 \pi G_4}\biggl({1\over r_h}+{1\over 3 r_{h}^3}-{q^2\over 15 r_{h}^5}  \biggr) \nonumber\\
{\omega_2 q^2 \gamma^2 \over\pi G_4}\biggl({1495 q^4\over 1512 r_{h}^9} -{421 q^2\over 504 r_{h}^7}-{3781 q^2\over 804 r_{h}^5}+{8\over 21 r_{h}^5}
+{80\over 21 r_{h}^3}+{24\over 7 r_{h}} \biggr)
\label{Gibbs2}
\end{eqnarray}
\begin{eqnarray}
F= {\omega_2\over 16 \pi G_4}\biggl(r_h-r_{h}^3+{3q^2\over r_{h}} \biggr)+{\omega_2 q^2 \gamma \over 8 \pi G_4}\biggl({5\over r_h}+{13\over 3 r_{h}^3}-{11q^2\over 5 r_{h}^5}  \biggr) \nonumber\\
{\omega_2 q^2 \gamma^2 \over\pi G_4}\biggl({691 q^4\over 1512 r_{h}^9} -{1185 q^2\over 504 r_{h}^7}-{313 q^2\over 40 r_{h}^5}+{8\over 3 r_{h}^5}
+{176\over 21 r_{h}^3}+{40\over 7 r_{h}} \biggr)
\label{Helmholtz2}
\end{eqnarray}
%%%%%%%%%%%%%%%%%%%%%%%%%%%%%%  
\begin{table}
\centering
\makebox[0pt][c]{\parbox{1.2\textwidth}{%
\begin{minipage}[b]{0.32\hsize}\centering
\begin{tabular}{ | c | c |c |}
\multicolumn{3}{c}
{$q=1/6$, $\gamma=0.001$} \\
\hline
$r_h$ & $\omega$ (with $\gamma$) & $\omega$ (with $\gamma^2$) \\ 
\hline
0.6 & 2.38326 - 1.61749 I & 2.38334 - 1.61749 I  \\ 
0.8 & 2.56044 - 2.15069 I & 2.56047 - 2.15069 I \\
1.0 & 2.78126 - 2.68328 I & 2.78128 - 2.68328 I \\ 
1.5 & 3.45627 - 4.01266 I & 3.45628 - 4.01266 I \\ 
2.0 & 4.23132 - 5.34181 I & 4.23132 - 5.34181 I \\
3.0 & 5.91476 - 8.00177 I & 5.91476 - 8.00177 I \\ 
\hline
\end{tabular}
\end{minipage}
\hspace{3cm}
\begin{minipage}[b]{0.32\hsize}\centering
\begin{tabular}{ | c | c | c |}
\multicolumn{3}{c}
{$q=1/20$, $\gamma=0.002$} \\
\hline
$r_h$ & $\omega$ (with $\gamma$) & $\omega$ (with $\gamma^2$) \\
\hline
0.6 & 2.42728 - 1.58306 I & 2.42731 - 1.58305 I  \\ 
0.8 & 2.58534 - 2.13225 I & 2.58535 - 2.13225 I \\
1.0 & 2.79670 - 2.67230 I & 2.79671 - 2.67230 I \\ 
1.5 & 3.46219 - 4.00870 I & 3.46219 - 4.00870 I \\ 
2.0 & 4.23412 - 5.33999 I & 4.23412 - 5.33999 I \\
3.0 & 5.91567 - 8.00119 I & 5.91567 - 8.00119 I \\ 
\hline
\end{tabular}
\end{minipage}
}}
\caption{\small Comparison between quasinormal modes at the order $\gamma$ with order $\gamma^2$. The left table corresponds to ($q=1/6$, $\gamma=0.001$) and the right 
table corresponds ($q=1/20$, $\gamma=0.002$). } 
\label{QNMcomparison}
\end{table}
%%%%%%%%%%%%%%%%%%%%%%%%%%%%%%  
%%%%%%%%%%%%%%%%%%%%%%%%%%%%%% 
We can again see that $G=S_{Total}/\beta$ even at $\gamma^2$ order. It implies that the first law thermodynamics is also satisfied at $\gamma^2$ order for the Weyl corrected 
black hole geometry. In figs.(\ref{FVsTgamma1by1000comparison}) and (\ref{FVsTgamma1by500comparison}), Helmholtz free energy at order $\gamma$ and at order $\gamma^2$ are plotted. We see that corrections due to 
$\gamma^2$ order are indeed very small, which again crudely justifies our choice of ``smallness'' of the values of $\gamma$ and the validity of the linear order expansion 
used in our calculations.
\\
\\
In a similar manner, QNMs at order $\gamma^2$ show very small change in magnitude compared to the QNMs at order $\gamma$. This is shown in Table (\ref{QNMcomparison}), where we see that
change in the QNM value occurs at fourth or fifth decimal place when $\gamma^2$ correction is taken into account.

\end{document}